\documentclass[pdflatex,sn-mathphys-ay]{sn-jnl}

\usepackage{graphicx}%
\usepackage{multirow}%
\usepackage{amsmath,amssymb,amsfonts}%
\usepackage{amsthm}%
\usepackage{mathrsfs}%
\usepackage[title]{appendix}%
\usepackage{xcolor}%
\usepackage{textcomp}%
\usepackage{manyfoot}%
\usepackage{booktabs}%
\usepackage{algorithm}%
\usepackage{algorithmicx}%
\usepackage{algpseudocode}%
\usepackage{listings}%
\usepackage{bm}
\usepackage{tikz}
\usepackage{lipsum}
\usepackage{dsfont}
\usepackage{amssymb}
\usepackage{physics}

\theoremstyle{thmstyleone}%
\theoremstyle{thmstyletwo}%

\theoremstyle{thmstylethree}%

\usepackage{etoolbox} 

\begin{document}

\title[High-expressibility Quantum Neural Networks using only
classical resources]{High-expressibility Quantum Neural Networks using only
classical resources}


\author*[1]{\fnm{Marco} \sur{Maronese}}
\email{marco.maronese@leonardo.com}
\author[1]{\fnm{Francesco} \sur{Ferrari}}
\author[1]{\fnm{Matteo} \sur{Vandelli}}
\author[1,2]{\fnm{Daniele} \sur{Dragoni}}

\affil[1]{\orgdiv{Quantum Computing Solutions}, \orgname{Leonardo S.p.A.}, \orgaddress{\street{Via R. Pieragostini 80}, \city{Genova}, \postcode{16151}, \country{Italy}}}

\affil[2]{\orgdiv{Hypercomputing Continuum Unit}, \orgname{Leonardo S.p.A.}, \orgaddress{\street{Via R. Pieragostini 80}, \city{Genova}, \postcode{16151}, \country{Italy}}}

\abstract{
Quantum neural networks (QNNs), as currently formulated, are near-term quantum machine learning architectures that leverage parameterized quantum circuits with the aim of improving upon the performance of their classical counterparts. 
In this work, we show that some desired properties attributed to these models can be efficiently reproduced without necessarily resorting to quantum hardware. We indeed study the expressibility of parametrized quantum circuit commonly used in QNN applications and contrast it to those of two classes of states that can be efficiently simulated classically: matrix-product states (MPS), and Clifford-enhanced MPS (CMPS), obtained by applying a set of Clifford gates to MPS. In addition to expressibility, we assess the level of primary quantum resources, entanglement and non-stabilizerness (a.k.a. "magic"), in random ensembles of such quantum states, tracking their convergence towards the Haar distribution. While MPS require a large number of parameters to effectively reproduce an arbitrary quantum state, we find that CMPS approach the Haar distribution more rapidly, in terms of both entanglement and magic. Our results on states with up to 20 qubits indicate that high expressibility in QNNs is attainable with purely classical resources.
}


\maketitle

\section{Introduction}

Quantum computation leverages the capability of qubits to encode exponentially more information than the same number of classical bits~\citep{nielsen2010quantum}.
This property motivated the development of quantum machine learning (QML) models~\citep{biamonte2017quantum,schuld2021machine,cerezo22challenges}, which exploit the exponentially large Hilbert space of qubits as effective latent space~\citep{schuld2019quantum, lloyd2020quantum, ahmadi2024quantifying}. Specifically, Quantum Neural Networks (QNNs) are learning models that mimic the functionality of classical neural networks within a quantum computational framework \citep{benedetti2019parameterized, schuld2014quest, ciliberto2018quantum, maronese2021continuous, maronese2022quantum}. QNNs are typically composed of three stages: embedding of the input data in the Hilbert space of the qubits; processing of the quantum state by a parametrized quantum circuit (PQC), namely a unitary operator $U(\bm \theta)$; finally, information extraction by measurements of certain observables of choice~\citep{schuld2020circuit}. QNN models have been proposed as candidates to enhance expressivity compared to classical neural networks, e.g. for classification or generative tasks~\citep{du2020expressive,abbas2021power}.

In analogy to the universal approximation theorem of classical statistical learning theory~\citep{vapnik1999nature, vapnik1999overview, de2018statistical}, a desirable property of QNN architectures is the ability to approximate a wide range of functions~\citep{schuld2021effect,gonon2025universal}. This property is closely related to the \emph{expressibility} of the underlying PQC, which is the capability to uniformly represent any pure quantum state in the Hilbert space ~\citep{sim2019expressibility,hubregtsen2021evaluation}. The latter task requires a number of parameters that grows exponentially with the number of qubits~\citep{nielsen2010quantum,harrow2009efficient}. However, in practical applications, PQCs should be designed to allow efficient implementation on actual quantum hardware, using a manageable number of one- and two-qubit operations.
This raises the question of which classes of PQCs are best suited to effectively span the Hilbert space. 

\begin{figure}[t]

\centering \includegraphics[width=.6\textwidth]{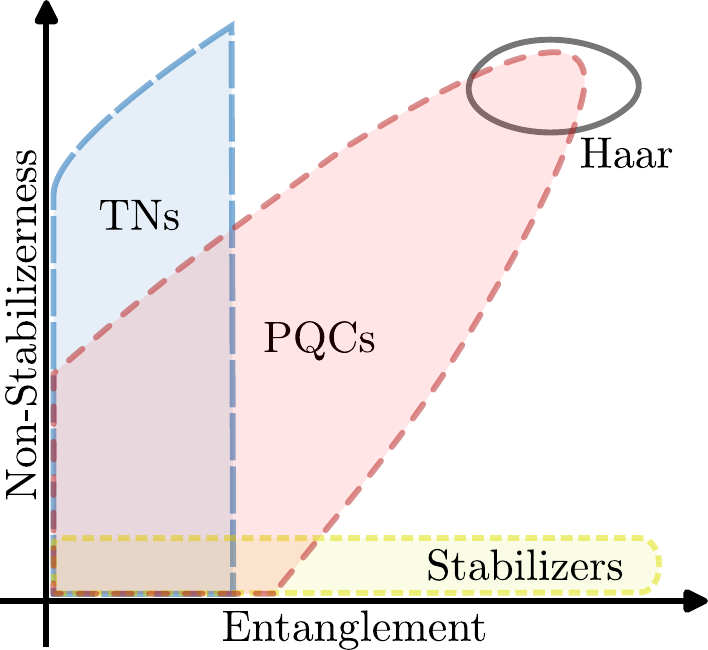}
  \caption{Qualitative representation of the phase space defined by non-stabilizerness (magic) and entanglement, illustrating the regions typically explored by different classes of wave function Ansätze. Tensor networks (TNs), shown in blue, can access areas with relatively low entanglement, but potentially high magic~\citep{chen2024magic_mps}. Parameterized quantum circuits (PQCs), depicted in red, generally have a layered structure that allows them to simultaneously increase both magic and entanglement, when the circuit is made deeper. They tend to reach the quantum resources of the Haar distribution (region bounded in gray) more efficiently than TNs. Stabilizer states, shown in yellow, have zero magic.}
  \label{fig:magic_vs_ent_qual}
\end{figure}

Previous works have explored the connection between expressibility of representative PQC architectures and their degree of quantumness, as measured for instance by entanglement~\citep{sim2019expressibility, ballarin2023entanglement}. Recently, a second measure of non-classicality of quantum states has emerged: non-stabilizerness, also known as \textit{magic}, which quantifies the amount of non-Clifford gates involved in a PQC~\citep{gottesman1998heisenberg,nielsen2010quantum,veitch2014resource}. Both of these resources, entanglement and magic, are deeply linked to the ability to efficiently emulate quantum states on a classical computer. Specifically, states with low entanglement can be efficiently simulated using tensor-network (TN) methods, e.g. by matrix-product states (MPS)~\citep{schollwoeck2011thedensity}, regardless of their level of magic (see Fig. \ref{fig:magic_vs_ent_qual}). Conversely, zero-magic states, commonly known as stabilizer states, can be simulated in polynomial time by leveraging the Gottesman-Knill theorem independently of their entanglement~\citep{gottesman1998heisenberg, aaronson2004improved}. Given that randomly sampled states typically exhibit large values of magic and entanglement~\citep{szombathy2025independent, iannotti2025entanglement}, it is an open question whether expressibility is ultimately rooted in both these non-classical resources. To this end, we characterize the quantum resource content and the expressibility of three different classes of quantum states. On the one hand, we consider states obtained by a prototypical quantum circuit for QML applications that is classically hard to emulate but runs on quantum hardware with polynomial resources. We refer to this architecture as fully-quantum neural network (fQNN). On the other hand, we consider quantum states that can be efficiently simulated classically: MPS and Clifford-enhanced matrix-product states (CMPS). The latter are obtained by applying a Clifford circuit to an MPS and exhibit large entanglement along with substantial magic. Expectation values over these states can be efficiently computed on a classical computer~\citep{lami2025quantum}. We generate random samples from the different classes of states and analyze the properties of these ensembles compared to uniformly distributed states as drawn from the Haar measure~\citep{mele2024introduction}. Our findings demonstrate that, using only polynomial classical resources, CMPS are able to compete with quantum native architectures like fQNN in terms of expressibility and quantum resources.

\section{Theory}

\subsection{Expressibility of quantum neural networks}

QNNs can be defined as parametrized learning models in the form:
\begin{equation}
    f(\bm{x}, \bm{\theta}) = \mathrm{Tr}\hspace{1pt}\left[OU(\bm{\theta})\overline{\rho}(\bm{x}) U^{\dag}(\bm{\theta})\right], 
    \label{eq:1}
\end{equation}
where $\bm{x}$ indicates the input data and $\bm{\theta}$ are the model parameters.
Such a model is to be interpreted as a composition of operations: 
\begin{enumerate}

   \item \textit{Embedding layer} (often referred to as feature map): the input $\bm{x}$ is encoded into the quantum state $\ket{\varphi(\bm{x})}$ in a $n$-qubit Hilbert space $\mathcal{H}_n$ with dimension $d=2^n$ \citep{schuld2019quantum}. Its density matrix is $\overline{\rho}(\bm{x}) = \ket{\varphi(\bm{x})}\bra{\varphi(\bm{x})}$;

   \item \textit{Parametrized unitary}: a unitary operator $U(\bm{\theta})$ acting on the states in $\mathcal{H}_n$, where $\bm{\theta}$ is a vector of $\cal P$ real parameters. We denote by $\Omega \in R^{\cal P}$ the domain of the parameters.

   \item \textit{Activation function}: a measurement-induced activation function $\rho \mapsto \mathrm{Tr}\hspace{1pt}\left[O \hspace{1pt}\rho \right]$ that extracts information from the Hilbert space by measuring an observable $O$ on the state with density-matrix $\rho$.
\end{enumerate}
The learning performance of the QNN model naturally depends on all three stages. However, for a fixed embedding layer, the capability of the QNN to effectively explore the quantum Hilbert space depends on the form of the parametrized unitary and is tied to the notion of \textit{expressibility}, which quantifies how well $U(\bm{\theta})$ can generate a distribution of states which are uniformly distributed across the Hilbert space.

To rigorously define this concept, we utilize the machinery of the Haar measure~\citep{mele2024introduction}, which identifies the unique left- and right-invariant probability measure on the group of unitary operators $\mathrm{U}(d)$. Applying unitary operators drawn from the Haar measure to a certain reference state\footnote{The choice of the reference state is arbitrary and does not lead to a loss of generality for the present discussion.}, we obtain the uniform distribution of quantum states across the Hilbert space, henceforth denoted as $\mu_\text{H}$. Similarly, we introduce the distributions of states $\mu$ obtained by the parametrized unitary operator of a chosen QNN model, namely the distribution of states ${\ket{\psi(\bm{\theta})}=U(\bm{\theta})\ket{0}^{\otimes n}}$ when varying the parameters $\bm{\theta}$. The ability of the distribution $\mu$ to approximate $\mu_\text{H}$ quantifies the expressibility of $U(\bm{\theta})$. We aim at quantifying the distance between $\mu$ and $\mu_\text{H}$ by examining the discrepancy between the moments of the two distributions.

To this aim, we introduce the quantity 
\begin{equation}
    A_t = \underset{\ket{\psi} \sim \mu}{\mathbb{E}} \left[\ketbra{\psi}^{\otimes t}\right] - \underset{\ket{\psi} \sim \mu_\text{H}}{\mathbb{E}} \left[\ketbra{\psi}^{\otimes t}\right] 
    \label{eq:t-state},
\end{equation}
where $\mathbb{E}[\cdot]$ denotes the expectation value over a distribution of quantum states. $A_t$ quantifies the difference between the $t$-th moment of $\mu$ and $\mu_\text{H}$. The distribution $\mu$ is said to form a $t$-state design if the statistical moments up to order $t$ match those of the uniform distribution $\mu_\text{H}$, i.e. if $A_t=~\bm{0}$~\citep{sim2019expressibility}. 
In practical terms, parametrized states ${\ket{\psi(\bm{\theta})}}$ are generated by sampling the parameters $\bm{\theta}$ from a distribution $\eta _{\theta}$ defined over the parameter space $\Omega$.
The expression in Eq.~\eqref{eq:t-state} for the matrix $A_t$ effectively translates to
\begin{equation}
A_t = \underset{{\bm \theta} \sim \eta_\theta}{\mathbb{E}} \left[\ketbra{\psi(\bm \theta)}^{\otimes t}\right] - \underset{\ket{\psi} \sim \mu_\text{H}}{\mathbb{E}} \left[\ketbra{\psi}^{\otimes t}\right].
\label{eq:A_matrix}
\end{equation}
where the expectation value $\mathbb{E}_{\bm{\theta} \sim \eta_\theta}[\cdot]$ is over the distributions of parameters $\eta _{\theta}$ defining the quantum state $\ket{\psi(\bm \theta)}$. 
To represent the difference in Eq.~\eqref{eq:A_matrix} by a single scalar value, we evaluate the Hilbert-Schmidt norm of $A_t$
\begin{equation}  
    \|A_t\|^2_{HS} := \mathrm{Tr}\left[A_t^{\dagger} A_t\right] = \mathcal{F}^{(t)}_{\mu} - \mathcal{F}^{(t)}_{\text{H}}, 
    \label{eq:diff_frame}
\end{equation}
where $\mathcal{F}^{(t)}_{\mu}$ is the so-called $t$-frame potential of the $\mu$ distribution and is defined as 
\begin{align}
    \mathcal{F}^{(t)}_{\mu} := \underset{\substack{{{\bm \theta} \sim \eta_\theta}\\{\bm \phi} \sim \eta_\phi}}{\mathbb{E}} \left[|\langle {\psi(\bm{\theta})}|{\psi(\bm{\phi})}\rangle|^{2t} \right].
    \label{eq:frame_potential}
\end{align}
The derivation of Eq.~\eqref{eq:diff_frame} from Eq.~\eqref{eq:A_matrix} is shown in~\citep{mele2024introduction}. The frame potentials for the Haar distributions are analytically known $\mathcal{F}^{(t)}_\text{H} = \frac{t! (d-1)!}{(d-1+t)!}$. Frame potentials can be interpreted as the moments of the distribution of fidelities $P_{\mu}(F=|\langle {\psi(\bm{\theta})}|{\psi(\bm{\phi})}\rangle|^{2})$ between pairs of states extracted from the same distribution $\mu$. Indeed, $\mathcal{F}_\mu^{(t)} = \mathbb{E}_{F\sim P_\mu}[F^t]$.
For what concerns the Haar distribution, the analytical expression of the distribution of fidelities is known to be ${P_{\text{H}}(F) = (d-1)(1-F)^{d-2}}$~\citep{mele2024introduction}.

We note that, since $\|A\|^2_{HS}$ is positive semidefinite, the condition $\mathcal{F}^{(t)}_{\mu} \geq \mathcal{F}^{(t)}_\text{H}$ strictly holds for any distribution of states $\mu$.
For this reason, $\mathcal{F}^{(t)}_\text{H}$ is a lower bound (called the \textit{Welch bound}). If the equality $\mathcal{F}^{(t)}_{\mu} = \mathcal{F}^{(t)}_\text{H}$ holds for each $t^* \leq t$, then $\mu$ is a $t$-state design. The equality holds for all $t$ if and only if $\mu$ is the Haar distribution.
In this framework, we can estimate expressibility by computing the first $t$ frame potentials, and thus determining if the distribution $\mu$ is a $t$-state design.
In circuit-based quantum computation models, the question has been raised whether it is possible to approximately saturate the Welch bound using PQCs with a polynomial gate scaling $O(\text{poly}(n))$. To approximate a $t$-design, the depth of a PQC must grow with both the number of qubits \(n\) and the order \(t\) as $O(\text{poly}(n, t))$ \citep{harrow2009random}. For many practical applications, logarithmic or polynomial scaling in \(n\) is sufficient for low-order designs (\(t \leq 3\)).

A more holistic approach to quantifying the mismatch between \( \mu \) and the Haar distribution involves directly estimating the distance between the corresponding fidelity distributions \( P_{\mu}(F) \) and \( P_{\text{H}}(F) \). One suitable metric for this purpose is the Kullback–Leibler divergence, which provides a quantitative measure of the difference between these distributions
\begin{align}
D_{\rm KL}(P_{\mu} &\parallel P_{\text{H}}) = \sum_F P_{\mu}(F) \log \left(\frac{P_{\mu}(F)}{P_{\text{H}}(F)}\right).
\label{eq:expressibility}
\end{align}
The smaller the value of $D_{\rm KL}$, the closer the distribution of fidelities between states sampled by the distribution $\mu$ is to that of the Haar-random states, indicating higher expressibility~\citep{sim2019expressibility}.  Recent work suggests that this expressibility metric is substantially correlated with the classification accuracy achieved by QNNs~\citep{hubregtsen2021evaluation}.

\subsection{Quantum resources: Entanglement and magic}

Alongside expressibility, we seek to characterize the non-classicality inherent in the distribution of parameterized quantum states $\mu$. To this end, we quantitatively evaluate two fundamental quantum resources: entanglement and magic.

To quantify the degree of entanglement in a quantum state, we use the (bipartite) entanglement entropy. This is defined by first partitioning the state into two subsystems. To do this, we label the qubits with sequential integer indices ${l=1,\dots,n}$. We then consider the various bipartitions $\{A_l,B_l\}_{l=1,\dots,n-1}$ that cut the system between the $l$-th and $(l+1)$-th qubit, defined by $A_l=[1\mathord{:}l]$ and $B_l=[l+1\mathord{:}n]$. The entanglement entropy is taken to be the maximum entropy over all such bipartitions~\citep{ballarin2023entanglement}
\begin{equation}\label{eq:ent_entropy}
\mathcal{S}_n=\max_{\{A_l,B_l\}} \Tr_{A_l}[\rho_{A_l} \log_2
 \rho_{A_l}],
\end{equation}
where $\rho_{A_l}=\Tr_{B_l}[\rho(\bm{\theta})]$ with
$\rho(\bm{\theta})=\ketbra{\psi(\bm{\theta})}$. 

Turning to magic, we utilize the stabilizer R\'enyi entropy as a measure of non-stabilizerness of a given quantum state~\citep{leone2022stabilizer}. To introduce this quantity, we briefly review the theory of stabilizers~\citep{gottesman1997stabilizer,gottesman1998heisenberg}. We begin by defining the set of Pauli strings for a system of $n$ qubits as
\begin{equation}
P_n = \{\zeta \; \sigma_1 \otimes \sigma_2 \otimes \cdots \otimes \sigma_n \}.
\end{equation}
with global phase $\zeta \in \{\pm 1, \pm i\}$ and Pauli matrices ${\sigma_l\in \{\mathds{1},X,Y,Z\}}$. A quantum state $\ket{\psi}$ is called a \textit{stabilizer state} if there exist an Abelian subgroup $\Sigma \subset P_n$ containing $d$ Pauli strings such that $s \ket{\psi}=\ket{\psi}, \ \forall s \in \Sigma$. 
Stabilizer states are transformed into other stabilizer states under the application of a special class of unitary operators, forming the so-called Clifford group, which is the normalizer group of $P_n$. Clifford operators are those that can be constructed by a combination of only Hadamard, $S$ and CNOT gates. In this respect, any stabilizer state can be obtained by applying a certain Clifford operator to the reference state $\ket{0}^{\otimes n}$. Following the Gottesman–Knill theorem, any quantum circuit composed solely of stabilizer state preparations, Clifford gates and Pauli measurements can be simulated in polynomial time on a classical computer~\citep{gottesman1998heisenberg, aaronson2004improved}. For this reason, quantifying the amount of non-stabilizerness (i.e., magic) of a quantum state is considered a way to assess its degree of non-classicality. Indeed, stabilizer states and Clifford gates are not sufficient for universal quantum computation~\citep{nielsen2010quantum}.

Among the different metrics for magic~\citep{veitch2014resource,beverland2020lower,hahn2022quantifying,haug2023stabilizerentropies,haug2023scalable,turkeshi2023measuring,ahmadi2024quantifying}, we select the stabilizer 2-R\'enyi entropy~\citep{leone2022stabilizer,leone2024stabilizerentropies}
\begin{equation}\label{eq:magic_entropy}
 \mathcal{M}_n= -\log_2 \sum_{P \in P_n} \Xi_P^2 - n
\end{equation}
where $\Xi_P=d^{-1}|\langle \psi(\bm{\theta}) | P | \psi(\bm{\theta}) \rangle|^2$. This quantity is zero if and only if the quantum state is a stabilizer and is additive with respect to the tensor product of states. Most importantly, it is invariant under the application of Clifford operators to the state $\ket{\psi(\bm{\theta})}$.

To simplify the following discussion, we introduce normalized entanglement and magic entropy with respect to the values obtained for the Haar distribution, i.e.
\begin{align}
    {\widetilde{\mathcal{M}}_{n}} = \frac{ \mathcal{M}_{n} }{\underset{\ket{\psi} \sim \mu_{\text{H}}}{\mathbb{E}}[\mathcal{M}_{n}]},
    \quad {\widetilde{\mathcal{S}}_{n}} = \frac{ \mathcal{S}_{n} }{\underset{\ket{\psi} \sim \mu_{\text{H}}}{\mathbb{E}}[\mathcal{S}_{n}]}
\end{align}
As the number of qubits increases, the entanglement and non-stabilizerness of a Haar-distributed state ensemble quickly become extremely localized around the asymptotic values
$\mathcal{M}_{n}= n-2$ and $\mathcal{S}_{n}= n/2$, respectively~\citep{szombathy2025independent, iannotti2025entanglement,turkeshi2025pauli}. The intensive quantities ${\widetilde{\mathcal{M}}_{n}}$ and ${\widetilde{\mathcal{S}}_{n}}$ indicate how close magic and entanglement of a given circuit are to the asymptotic values of the Haar ensemble.

\subsection{Classically-hard \emph{vs} classically simulatable QNN models}

\begin{figure*}[t]
  \centering \includegraphics[width=1\textwidth]{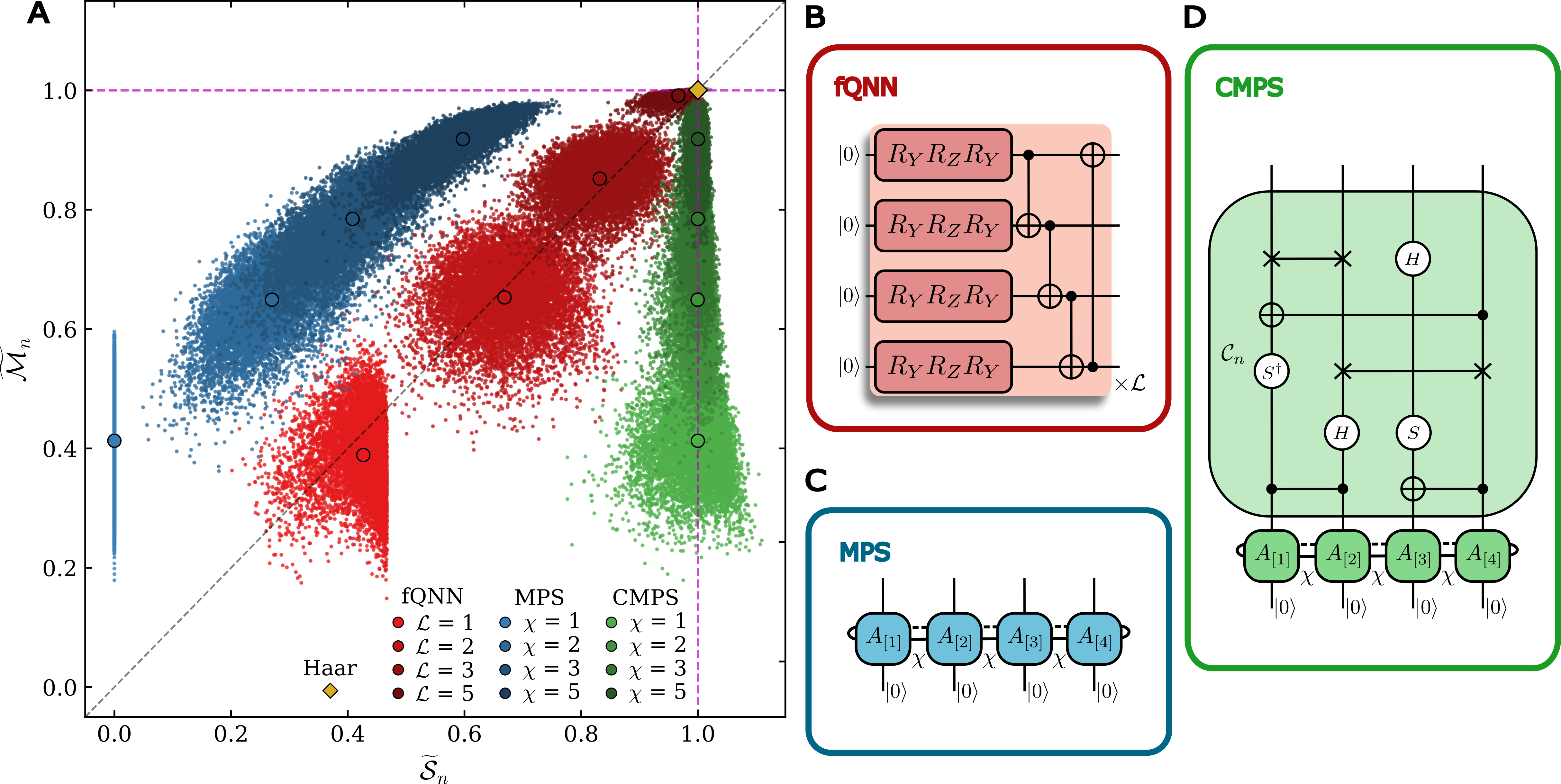}
  \caption{\textbf{A} Quantum states samples in the entanglement-magic phase space ${\widetilde{\mathcal{S}}}-{\widetilde{\mathcal{M}}}$. The results are obtained for $n=10$ qubits. States sampled from the Haar measure form an extremely localized distribution~\citep{szombathy2025independent} centered at $(1,1)$ (marked by the gold diamond). The other colors correspond to the various architectures shown in \textbf{B},\textbf{C},\textbf{D}. Red colors indicate fQNN states (\textbf{B}) with different number of layers, blue colors indicate MPS states (\textbf{C}) with different bond dimension, while green colors denote CMPS (\textbf{D}) results. Empty circles correspond to the average values of ${\widetilde{\mathcal{S}}}$ and ${\widetilde{\mathcal{M}}}$ for the various sets of samples.}
  \label{fig:magic_vs_ent}
\end{figure*}

In this work, we consider three different types of states, which can be used as QNN architectures.

\emph{fQNN}: a widely used PQC for QML. The name stands for fully-quantum neural network. The network is characterized by a number $\mathcal{L}$ of layers. Each layer is defined by generic $SU(2)$ rotations applied independently to each qubit and decomposed using the Euler angles decomposition $R_Y(\alpha)R_Z(\beta)R_Y(\gamma)$. These operations are followed by CNOT gates arranged in a ring topology, as sketched in Fig.~\ref{fig:magic_vs_ent}\textbf{B}. These states cannot be represented efficiently on a classical computer, but can be executed on quantum hardware with polynomial resources. An fQNN with $\mathcal{L}$ layers is specified by a total of $\mathcal{P}=3n\mathcal{L}$ angles as parameters. We generate random fQNN by indepently sampling each of these angles from the uniform distribution on the interval $\left[0, 2\pi\right)$.

\emph{MPS}: matrix-product states with periodic boundary conditions defined as
\begin{align}\label{eq:mps}
    |{\rm MPS}\rangle = \sum_{\{{\bf s}\}} \Tr \left[A^{s_1}_{[1]}  A^{s_2}_{[2]} \dots A^{s_n}_{[n]}\right] |s_1 \dots s_n\rangle \,,
\end{align}
where, for fixed $s_i \in \{0, 1\}$ (physical dimension), $A_{[i]}^{s_i}$ are $\chi \times \chi$ complex-valued matrices (as schematically illustrated in Fig.~\ref{fig:magic_vs_ent}\textbf{C}). The hyperparameter $\chi$ is conventionally called bond dimension. In principle, MPS can span the full Hilbert space $\mathcal{H}$ if equipped with an exponentially large bond dimension, i.e. if $\chi \sim 2^{n/2}$~\citep{eisert2013entanglement}. However, for practical application, polynomial bond dimensions $\chi = \mathrm{poly}(n)$ are employed. In this case, MPS can be efficiently represented and manipulated on a classical computer, allowing for tractable computation of expectation values of local observables, by means of tensor contraction techniques~\citep{schollwoeck2011thedensity,collura2024tensor}. In general, MPS provide an efficient representation for quantum states with limited (e.g., area-law) entanglement \citep{eisert2008area}. In the special case $\chi = 1$, $\ket{\rm MPS}$ reduces to a simple product state, meaning it has no entanglement. The MPS ansätze of Eq.~\eqref{eq:mps} requires a total of $\mathcal{P}=4n\chi^2$ real parameters, coming from its $2n$ $\chi \times \chi$ complex-valued matrices. 
We generate random MPS by independently sampling real and imaginary parts of each entry in the $A_{[i]}^{s_i}$ matrices from a normal distribution $\mathcal{N}(0,1)$~\citep{lancien2022correlation}.

\emph{CMPS}: states obtained by applying a Clifford operator to an MPS state (as exemplified in Fig.~\ref{fig:magic_vs_ent}\textbf{D}) \citep{mello2025clifford, masot2024stabilizer, mello2024hybrid, qian2024augmenting}. 
Explicitly, we can write
\begin{align}
    |{\rm CMPS}\rangle = {U}_C |{\rm MPS}\rangle.
\end{align}
Although the direct representation of these states has an exponential footprint on classical memory, we can efficiently compute expectation values of observables~\citep{lami2025quantum} required in QNN applications as 
\begin{equation}
            \langle {\rm CMPS}| O |{\rm CMPS}\rangle  
        = \langle {\rm MPS}| \left[{U}^\dagger_C\, O \, {U}^{\phantom{\dagger}}_C \right]|{\rm MPS}\rangle.
\end{equation}
The classical simulability of the CMPS ansatz relies on the specific structure of the expectation values required for variational tasks.
For an observable $O$ decomposed into a sum of $M$ Pauli strings, $O = \sum_{j=1}^{M} \alpha_j P_j$ the expectation value is computed as $\langle \mathrm{CMPS} \lvert O \rvert \mathrm{CMPS} \rangle
= \sum_{j=1}^{M} \alpha_j
\langle \mathrm{MPS} \lvert U_C^\dagger P_j U_C \rvert \mathrm{MPS} \rangle$.
The computation proceeds in two efficient steps. First, 
each Pauli string $P_j$ is transformed into a new Pauli string $P'_j = U_C^\dagger P_j U_C$ by Clifford conjugation,
using the symplectic tableau formalism~\citep{aaronson2004improved}. This operation scales classically as
$\mathcal{O}(n^2)$ per string.
In the following step, the expectation value $\langle \mathrm{MPS} \lvert P'_j \rvert \mathrm{MPS} \rangle$
is evaluated via tensor contraction. For an MPS with bond dimension $\chi$,
this contraction scales as $\mathcal{O}(n \chi^3)$.
Consequently, the total time complexity is polynomial in the system size,
scaling as $\mathcal{O}\!\left(M \left(n^2 + n \chi^3 \right)\right)$.
The memory footprint is similarly efficient, requiring $\mathcal{O}(n^2)$
to store the Clifford tableau and $\mathcal{O}(n \chi^2)$ for the MPS
parameters, ensuring the ansatz remains efficient.

Random sampling of CMPS states requires specifying $\mathcal{P}=4n\chi^2$ continuous parameters, which fully characterize the MPS part and are sampled as previously described. Additionally, the Clifford unitary $U_C$ needs to be chosen. Since the Clifford group for $n$ qubits is finite (and thus countable), each Clifford unitary can be uniquely defined by its action on the generators of the Pauli group, or equivalently, by a binary symplectic matrix, up to a global phase~\citep{hostens2005stabilizer}. In this framework, random Clifford circuits can be sampled by adopting the method of~\citep{bravyi2021logic}. On the contrary, we note that sticking to a single fixed Clifford unitary $\overline{U}_C$ is not suitable for our analysis, as it would yield the same values of $\mathcal{F}^{(t)}$ and $D_{\rm KL}$ as in the MPS case, as can be easily verified by substitution in Eqs.~\eqref{eq:frame_potential} and~\eqref{eq:expressibility}.

\section{Computational details}

Our calculations are based on the Pennylane Python software library for quantum computing~\citep{bergholm2018pennylane}, interfaced with JAX~\citep{jax2018github}. Random Clifford unitaries for the CMPS ensembles are generated using the \texttt{Tableau.random} subroutine of the Stim library~\citep{gidney2021stim}, which implements the method of~\citep{bravyi2021logic}, and the Cirq library~\citep{cirqcode}.
For the calculations presented in this work, we used a single node of our proprietary \textit{davinci-1} cluster equipped with 2 AMD EPYC Rome 7402 @ 2.80 GHz CPUs (24 cores each), 512 GB of RAM and 4 NVIDIA A100 GPUs with 40 GB of dedicated memory.

\section{Results}

\subsection{Magic \emph{vs} Entanglement}

We begin by assessing the quantum resource content of the three different kinds of states ensembles $\mu = \{$fQNN, MPS, CMPS$\}$, to understand where these are located in the entanglement-magic phase space. The results are shown in Fig.~\ref{fig:magic_vs_ent}\textbf{A}, for a system of $n=10$ qubits. Specifically, we present a scatter plot illustrating the distribution of 10,000 sample instances (for each circuit type) in the normalized entanglement and magic plane $\widetilde{\mathcal{S}}-\widetilde{\mathcal{M}}$, for different number of layers $\mathcal{L}$ (for fQNN) and different bond dimensions $\chi$ (for MPS, CMPS). We note that CMPS samples are obtained by taking the MPS states from the MPS samples and applying a random Clifford operator to them. Therefore, as expected, the magic content of the CMPS samples is equal to that of the MPS samples, due to the invariance of $\mathcal{M}_n$ under the application of a Clifford operator. As previously shown~\citep{szombathy2025independent}, Haar distributed states tend to form an extremely localized distribution in the entanglement-magic plane, which in our plot is centered around the saturation values $\widetilde{\mathcal{M}}_{\rm H}=1$ and $\widetilde{\mathcal{S}}_{\rm H}=1$.
The distribution of resources for all the $\mu$ state ensembles considered here localize more and more towards those of the Haar distribution, as their number of parameters $\mathcal{P}$ increases. Magic and entanglement of the different classes of states converge towards the values of the Haar distribution in different ways. For the fQNN case, we note that the distribution with $\mathcal{L}=1$ exhibits a sharp bound in entanglement because the circuit only has a single layer of CNOT gates. For $\mathcal{L} \geq 2$, instead, the convergence is roughly balanced in both resources, and the samples align along the diagonal of the plot. For what concerns MPS, $\chi=1$ corresponds to the case of simple product states, which have exactly zero entanglement. MPS samples approach the Haar limit from the upper-left direction of the ${\widetilde{\mathcal{S}}}-{\widetilde{\mathcal{M}}}$ plane, while CMPS states lie in the right-most side of the plane.

\begin{figure*}[t]

\includegraphics[width=1\textwidth]{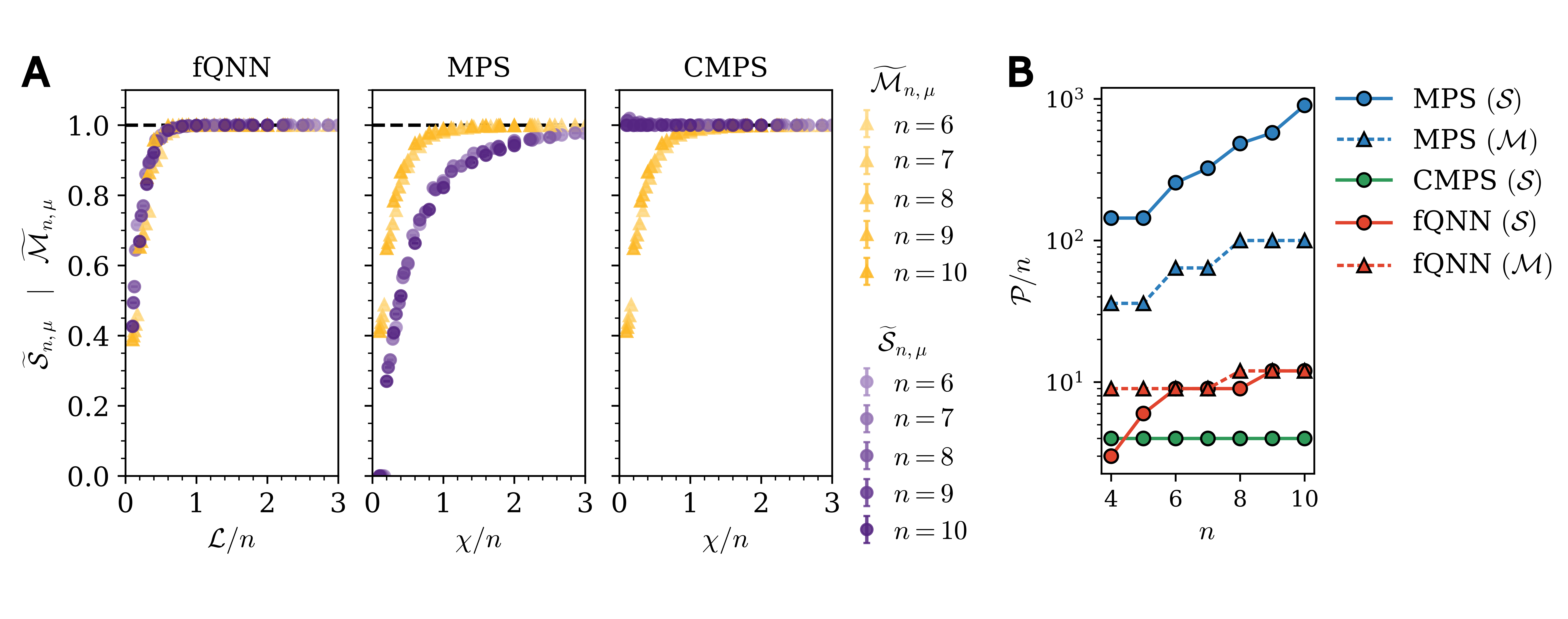}
  \caption{\textbf{A} Average values of $\widetilde{\mathcal{M}}$ (golden triangles) and $\widetilde{\mathcal{S}}$ (violet circles) for fQNN, MPS and CMPS ensembles, as a function of the number of layers $\mathcal{L}$ or the bond dimension $\chi$. Different shades of the colors indicate different numbers of qubits $n$. \textbf{B} Number of parameters $\mathcal{P}$ that are needed such that magic and entanglement of the various ensembles reach 90\% of the asymptotic values of the Haar distribution, namely  $\widetilde{\mathcal{M}}_{n, \mu}=0.9$ and $\widetilde{\mathcal{S}}_{n, \mu}=0.9$. Triangles and circles represent magic and entanglement, respectively. Different colors indicate different classes of states. Here $\widetilde{\mathcal{M}}_{n, {\rm CMPS}}$ is omitted since $\widetilde{\mathcal{M}}_{n, {\rm CMPS}} = \widetilde{\mathcal{M}}_{n, {\rm MPS}}$ by construction.}
  \label{fig:resource_speed}
\end{figure*}

To better illustrate the convergence of quantum resources distributions, in Fig.~\ref{fig:resource_speed}\textbf{A} we provide a detailed view of how their average values approach the asymptotic values of the Haar samples, for different number of qubits $n$. The average magic $\widetilde{\mathcal{M}}$ is shown as golden triangles and the entanglement $\widetilde{\mathcal{S}}$ as violet circles for each state type $\mu$. The fQNN resources are plotted in terms of the normalized number of layers $\mathcal{L} / n$, while the MPS and CMPS results are shown as a function of the normalized bond dimension $\chi / n$. With this choice, the results are independent on $n$ and the data points of different system sizes fall on the same curve.

We observe that fQNN states converge in both resources at the same rate with increasing layers, as the points for $\widetilde{\mathcal{M}}_{\rm fQNN}$ and $\widetilde{\mathcal{S}}_{\rm fQNN}$ overlap when plotted against $\mathcal{L}/n$ (see Fig.~\ref{fig:resource_speed}\textbf{A}, left-most panel). On the other hand, the MPS states saturate rather quickly in terms of magic but their entanglement converges slowly as the number of parameters increases, compared to the other states. Finally, we find that the CMPS displays a peculiar behavior. Indeed, the mean entanglement reaches an entanglement content compatible with the Haar asymptotic value already at $\chi=1$. From this perspective, increasing $\chi$ only improves the content of magic $\widetilde{\mathcal{M}}_{\rm CMPS}$. Since $\widetilde{\mathcal{S}}_{\rm CMPS}$ is always compatible with $\widetilde{\mathcal{S}}_{\rm H}=1$ and $\widetilde{\mathcal{M}}_{\rm CMPS}$ saturates quickly, we observe that the CMPS contains a high degree of quantumness despite being classically simulatable for the purpose of QNNs.

With an eye on QNN applications, in Fig.~\ref{fig:resource_speed}\textbf{B}, we discuss the normalized number of parameters $\mathcal{P}/n$ necessary to reach the magic and entanglement values of the Haar distribution. As a criterion, we plot at which value of $\mathcal{P}/n$ the various classes of states reach 90\% of the resources of the Haar-distribution, in analogy with the definition of \emph{entangling layers} used in~\citep{ballarin2023entanglement}. Although limited to a small range of qubits $n$, we can qualitatively estimate the scaling of $\mathcal{P}/n$ for the various resources. Concerning $\widetilde{\mathcal{M}}$, we observe that the scaling behaviors as a function of $n$ appear to be compatible with a polynomial scaling for all $\mu$. This observation also applies to MPS and CMPS. On the other hand, for what concerns entanglement, we show that the value of $\mathcal{P}/n$ needed to reach $\widetilde{\mathcal{S}}=0.9$ for MPS is compatible with an exponential scaling as a function of $n$, in accordance with the known theoretical bound~\citep{PhysRevB.73.094423}. This means that an exponential $\mathcal{P}$ is needed to effectively reproduce the non-classicality of the Haar distribution as the feature space $\mathcal{H}_n$ increases in size. 

\begin{figure}[t]

 \centering \includegraphics[width=0.7\columnwidth]{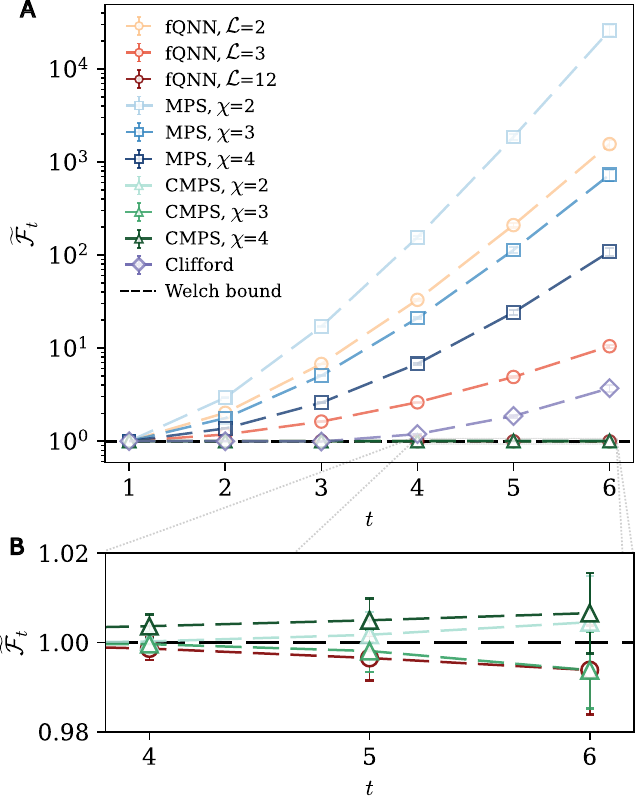}
  \caption{Rescaled frame potentials $\widetilde{\mathcal{F}}^{(t)}_\mu$ in function of the $t$, numerically estimated with $10^7$ samples of fidelities $F$. In panel \textbf{A} we show results for $n=10$ qubits and different classes of states. The Welch bound is marked by the black horizontal line.
  In panel \textbf{B} we show a zoom in a region close to the Welch bound for $t=4,5,6$.
  The results for CMPS are compatible with the Haar moments distribution even for a bond dimension of \(\chi = 2\). For a fair comparison, we show, for instance, the case of fQNN with \(\mathcal{L} = 12\) layers because it has the same number of parameters, \(\mathcal{P} = 360\), as a CMPS with bond dimension \(\chi = 3\). 
}
  \label{fig:welch_bound _saturation}
\end{figure}

\subsection{Frame potentials and Welch bounds}
\label{sec:welch}

As discussed previously, a way to estimate the deviations of a distribution of states from the Haar distribution is in terms of the frame potentials $\mathcal{F}^{(t)}_\mu$. 
We collect samples of fidelities $F=|\langle {\psi(\bm{\theta})}|{\psi(\bm{\phi})}\rangle|^2$ by generating $10^7$ pairs of states and we estimate the frame potentials as the moments of the distribution $P(F)$, according to Eq.~\eqref{eq:frame_potential}. For a given distribution $\mu$, all moments $\mathcal{F}^{(t)}_\mu$ are estimated using the same set of fidelity sample pools. Since the Welch bound $\mathcal{F}^{(t)}_{\rm H}$ is a lower bound of the frame potentials, we define rescaled potentials $\widetilde{\mathcal{F}}^{(t)}_\mu =\mathcal{F}^{(t)}_\mu/\mathcal{F}^{(t)}_{\rm H}$, which are computed for the different classes of states as a function of the order $t$. The lower-bound condition is then expressed in terms of the normalized potentials as $\widetilde{\mathcal{F}}^{(t)}_\mu \geq 1$, with $\widetilde{\mathcal{F}}^{(t)}_{\rm H} \equiv 1$.
In Fig.~\ref{fig:welch_bound _saturation}\textbf{A} we show the numerical estimation of the frame potentials for $n=10$ qubits as a function of the order $t$. In Fig.~\ref{fig:welch_bound _saturation}\textbf{B}, we provide a zoom of the results in the region around $\widetilde{\mathcal{F}}^{(t)} \approx 1$ for $t=4,5,6$ to resolve small differences from the Welch bound. All frame potentials are shown for selected values of \( \mathcal{L} \) (for fQNN) and \( \chi \) (for MPS and CMPS), which are chosen such that the total number of parameters \( \mathcal{P} \) of the different architectures is comparable.

We observe that fQNN with \( \mathcal{L} =2,3\) and MPS at low $\chi$ diverge super-exponentially from the Welch bound as a function of $t$. These state ensembles fail to reproduce moments of the Haar distribution beyond $t=1$.
However, fQNN with \( \mathcal{L} = 12\) (or $\mathcal{P} = 360$ parameters) outperforms the MPS both with \( \chi = 3\)  (same number of parameters) and \( \chi = 4\) ($\mathcal{P} = 640$).

In Fig.~\ref{fig:welch_bound _saturation}\textbf{B}, we highlight how the frame potentials $\widetilde{\mathcal{F}}^{(t)}$ of a fQNN with \( \mathcal{L} = 12\) are numerically compatible with the Welch bound within the error bars for all values of $t$ considered here. Similarly, also the frame potentials of CMPS agree with the Welch bound already for $\chi = 2$ within the same range of $t$. This suggests that CMPS require a smaller number of parameters $\mathcal{P}$ than fQNN to converge towards the Welch bound. For completeness, we also numerically demonstrate that stabilizer states generated by Clifford unitaries form a $3$-design, in agreement with known analytical results~\citep{webb2016theclifford,zhu2017multiqubit}.

It is important to note that, although we employ the Haar measure as a theoretical reference for assessing expressibility, our goal is not to demonstrate certified convergence to the Haar randomness. Indeed, establishing exact convergence from the frame potentials is infeasible, as numerical simulations are inevitably restricted to a finite order and theoretical results are available only for the lowest levels of $t$-design~\citep{lami2025quantum}. Rather, our focus is a comparative analysis between quantum-native architectures (fQNNs) and classically efficient models (CMPS).
Nevertheless, our results indicate that, at least for small system sizes, the low-order frame potentials of CMPS and fQNNs display comparable deviations from the Welch bounds and agree numerically within their error bars.

\subsection{Quantifying expressibility by Kullback-Leibler divergence}

\begin{figure}[t]
  \centering \includegraphics[width=0.8\columnwidth]{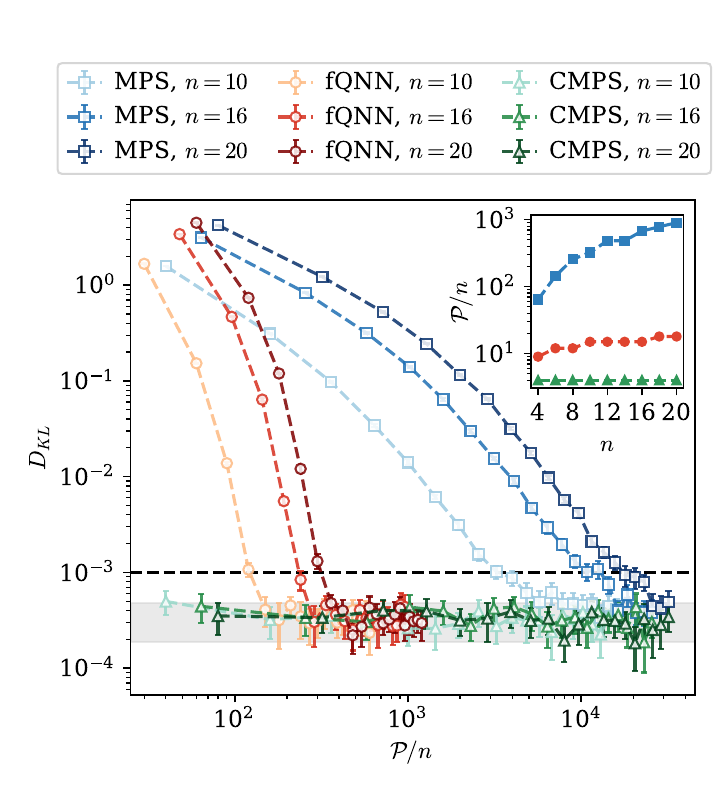}
  \caption{Kullback-Leibler divergence $D_{\rm KL}$ between numerically estimated fidelity distributions $\hat{P}_{\mu}(F)$, for different classes of states, and the exact Haar fidelity distribution $P_{\text{H}}(F)$. Error bars are computed using jackknife resampling. Lower $D_{\rm KL}$ divergence signifies higher expressibility of the architecture in approximating the Haar distribution on pure quantum states.
  The gray shaded region represents the KL divergence between the true Haar fidelity distribution $P_{\text{H}}(F)$ and the one obtained by sampling $10^5$ Haar-distributed fidelities for $n = 10,16,20$ qubits, accounting for the errors in this estimate.
  In the inset, the number of parameters, normalized by the number of qubits, required by different architectures to reach a threshold of $D_{\rm KL}=10^{-3}$ (black dashed line in the main plot) is shown. }
  \label{fig:exp}
\end{figure}

While frame potentials provide valuable insight into expressibility, their descriptive power is limited at low orders. To address this, we assess expressibility using a more global indicator based on the Kullback–Leibler divergence between the fidelity ditributions $P_{\mu}(F)$ and $P_{\rm H}(F)$ [Eq.~\eqref{eq:expressibility}]. This quantity is a rather simple but commonly used metric \citep{sim2019expressibility,rasmussen2020reducing,hubregtsen2021evaluation,nakaji2021expressibility,chen2021onthe,mohannad2022evaluation,aktar2023predicting,liu2025analysis,zhang2025learning}, which needs to be evaluated numerically in the case of an unknown distribution $P_{\mu}(F)$. To this end, we sample $10^5$ pairs of states and compute the corresponding fidelities. Creating a histogram from these samples to approximate the distribution $P_{\mu}(F)$ requires defining a binning strategy. For distributions with sharp peaks or rapid variations, selecting the bin width and number of bins is a particularly delicate task, as it strongly influences the accuracy and usefulness of the histogram as an approximation of the underlying distribution. As the number of qubits increases, the underlying distribution \( P_{\mu}(F) \) quickly becomes extremely peaked at \( F \sim 0 \), similar to the fidelities of Haar-distributed states \( P_{\text{H}}(F) \sim O(d(1-F)^{d}) \).
Therefore, instead of creating a histogram, we employ the Gaussian Kernel Density Estimation (KDE)~\citep{scott2015multivariate} method to approximate $P_{\mu}(F)$ from the samples of fidelities. KDE provides a smoother and more accurate representation of the distribution, especially for datasets with sharp peaks or rapid variations. We denote distributions numerically estimated by KDE with a hat, e.g. \( \hat{P}_\mu(F) \). The latter, together with the analytically known $P_{\rm H}(F)$, are used to compute the Kullback-Leibler divergence of Eq.~\eqref{eq:expressibility} for the different classes of states. The results are shown in Fig.~\ref{fig:exp}, for $n = 10,16,20$ qubits.

Knowing the analytical formula for $P_{\rm H}(F)$, we are able to assess the numerical error introduced by the finite sampling. This is achieved by computing the Kullback-Leibler divergence between the exact probability $P_{\rm H}(F)$ and the numerically estimated \( \hat{P}_{\rm H}(F) \), obtained by sampling $10^5$ pairs of Haar-distributed states. We obtain a discrepancy ${D_{\rm KL}(\hat{P}_{\rm H} \parallel P_{\rm H}) \approx 3 \times 10^{-4}}$ in the cases of $n = 10,16,20$ qubits, indicated by the shaded area in Fig.~\ref{fig:exp} (which accounts for error bars).
This means that \( D_{\rm KL}(\hat{P}_{\mu} \parallel P_{\text{H}}) \) computed from $10^5$ samples is approximately bounded from below by this region, regardless of the ensemble of states considered. This establishes the resolution limit of our numerical analysis, defining an empirical `zero' that characterizes the maximum convergence attainable with the chosen sample size. We observe how the values of $D_{\rm KL}$ of the different classes of states approach the lower bound as the number of parameters is increased. The results for fQNN and CMPS architectures converge significantly faster than MPS.

Considering expressibility as a relative metric that allows us to evaluate which architectures better approximate the Haar distribution, we choose a threshold value of $D_{\rm KL}=10^{-3}$ to assess the characteristic number of parameters needed to reach this level of approximation as the state size increases, and thus the number of qubits.
As shown in Fig.~\ref{fig:exp}, although fQNN is much faster than MPS in reaching the chosen bound, CMPS already surpasses the threshold for $\chi = 1$. We observe that in the range $n = 4$ to $n = 20$, a bond dimension $\chi = 1$ is always sufficient for CMPS, unlike the cases of fQNN and MPS where the number of parameters (normalized by the number of qubits) increases within this range of instances.

\section{Discussion}

\begin{figure}[t]
\centering \includegraphics[width=0.7\textwidth]{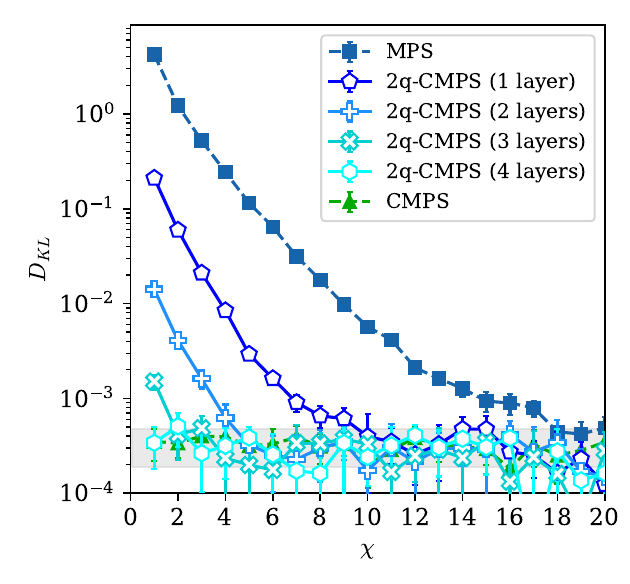}
\caption{Kullback-Leibler divergence $D_{\rm KL}$ between $\hat{P}_{\mu}(F)$ and the exact Haar fidelity distribution $P_{\text{H}}(F)$, analogous to Fig.~\ref{fig:exp}. Results for $n=20$ qubits and different values of the bond dimension $\chi$ are shown. Full squares (triangles) represent the data for MPS (CMPS) classes. Empty markers correspond to 2q-CMPS ans\"atze, a restricted class of CMPS in which the Clifford operator is constructed by repeated layers of cascaded two-qubit Clifford circuits~\citep{qian2024augmenting}.\label{fig:2q-CMPS} }
\end{figure}

In terms of quantum resources, the fQNN quantum circuit discussed in this work can be seen as a sequence of alternating layers that inject magic, via single-qubit rotations, and generate entanglement, through the ring of CNOT gates. In this respect, an isolated layer of the fQNN is equivalent to a CMPS with bond dimension 1, where the Clifford operator is the entangling map of CNOTs. The repeated application of these layers underscores the difficulty of classically simulating such architectures. In this paper we show that the high expressibility achieved by quantum circuits that alternate phases of magic and entanglement injection can still be obtained with CMPS architectures that are classically simulatable with polynomial computational cost and memory requirements. The key insight is that it suffices to partition the model so that the two resources, magic and entanglement, are injected sequentially, as is done in CMPS.
In particular, the process involves first using an MPS to inject magic, which grows rapidly with increasing bond dimension, and then applying a Clifford circuit to generate large entanglement (leaving the magic invariant). This approach potentially enables the construction of QNN with high expressibility, while retaining efficient classical simulability.

The high expressibility of CMPS does not identify which classes of quantum states are more efficiently approximated by this ansatz. In general, any quantum state can be represented as a CMPS given a sufficiently large bond dimension; however, in the worst-case scenario, this bond dimension scales exponentially with the number of qubits. We argue that CMPS could be more effective in representing quantum states characterized by low levels of non-stabilizerness entanglement entropy~\citep{huang2024nonstabilizerness}. This interpretation stems from the disentangling role of the Clifford operator~\citep{true2022transitions,frau2025}, which reduces the entanglement of the target state, lowering the bond dimension required for the underlying MPS. Future work could address the limits of expressibility of CMPS quantitatively investigating which states are harder to represent efficiently.

While our work is centered on the expressibility of the CMPS, their trainability remains an open and nontrivial problem. Broadly speaking, the training of generic CMPS requires the joint optimization of continuous MPS parameters and the discrete choice of the $n$-qubits Clifford circuit. However, variational optimization schemes for CMPS, used to target ground states of spin~\citep{qian2024augmenting,fan2025disentangling} and quantum chemistry models~\citep{fu2025clifford,huang2025augmenting}, were shown to be capable of consistently yielding better results than the corresponding bare MPS, with a relatively small computational overhead. The aforementioned works used a restricted ansatz for CMPS with layers of cascaded $2$-qubit Clifford circuits, hereafter referred to as 2q-CMPS, which are efficiently trainable by a Clifford augmented density matrix renormalization group (CA-DMRG) algorithm~\citep{qian2024augmenting}. This specific ansatz is less general than using $n$-qubits Clifford operators. Nevertheless, as shown in Fig.~\ref{fig:2q-CMPS} for a system $n=20$ qubits, the expressibility of 2q-CMPS converges towards generic CMPS when increasing the number of layers of 2-qubit Clifford circuits\footnote{Here, a single layer refers to a bidirectional cascade of 2-qubit Clifford circuits acting on adjacent sites, traversing the chain from left to right and returning in a reverse sweep. This mirrors the sweep structure of the CA-DMRG algorithm.}. Within the limitations of finite sampling, our results indicate that already few Clifford layers in 2q-CMPS are sufficient to reproduce the expressibility of the general CMPS ansatz.

\section{Conclusions}

In this work, we assessed the amount of magic and entanglement of three architectures for QNN applications: fQNN, MPS and CMPS. The first requires polynomial resources on a quantum computer, but is exponentially hard to emulate on classical hardware. In contrast, MPS and CMPS can be efficiently simulated on a classical computer. 
Our calculations show that CMPS display resource characteristics comparable to fQNN, using a similar number of parameters. In the context of QML, these findings challenge the assumption that states with both high entanglement and high magic are not efficiently accessible using only classical resources. 

To further investigate the connection between quantum resources and expressive power for QML, we assess how these architectures perform in terms of expressibility, namely the ability of a parameterized quantum state to approximate arbitrary random states. We find that both architectures achieve comparable levels of expressibility, according to common metrics based on the distribution of fidelities. This suggests that while fQNNs are promising for future quantum machine learning, CMPS can offer a practical, classically efficient alternative that requires no quantum hardware.

In addition to that, CMPS may also be useful in an integrated quantum-classical hybrid workflow, as they can be compiled in terms of quantum gates~\citep{ran2020encoding,malz2024preparation,smith2024constant}. This might enable training using only a classical computer~\citep{qian2024augmenting}, alleviating the burden of extensive measurements on a quantum hardware. 
Instead, the inference stage may be performed on quantum hardware, potentially leading to significant energy savings and speed improvements.

We note that high expressibility guarantees neither efficient trainability nor superior performance in machine learning applications; rather, high expressibility is often associated with optimization difficulties such as barren plateaus~\citep{holmes2022connecting}. In the case of CMPS, there is a further challenge related to the discrete choice of the Clifford operators, as outlined in the Discussion. Moreover, to date, CMPS have not been applied to standard machine learning tasks yet. Nevertheless, the feasibility of their optimization is strongly suggested by results in other domains.
In the context of solving hard many-body quantum problems, CMPS architectures have already been employed within recent advances in CA-DMRG and other hybrid stabilizer-tensor network variational schemes~\citep{qian2024augmenting, qian2024parent, fan2025disentangling, fu2025clifford}. The effectiveness of the CA-DMRG algorithm for 2q-CMPS~\citep{qian2024augmenting} may suggest that similar strategies can be effectively devised to train more generic Clifford ansätze.
Thus, the question of CMPS trainability remains an active area of research and poses challenges due to the coexistence of continuous and discrete degrees of freedom. Future work involves also the applications of these quantum-inspired methods to practical use cases, as they may help circumventing known issues of quantum on-chip training of QNN models.

\bmhead{Acknowledgements}
We acknowledge the whole Quantum Computing Solutions group at Leonardo S.p.A., in particular Francesco Turro, for insightful discussions. We are grateful to Damiano Paniccia for helping us with the JAX implementation. 
 
\bmhead{Funding Declaration}
This research received no external funding.



\begin{thebibliography}{79}
\ifx \bisbn   \undefined \def \bisbn  #1{ISBN #1}\fi
\ifx \binits  \undefined \def \binits#1{#1}\fi
\ifx \bauthor  \undefined \def \bauthor#1{#1}\fi
\ifx \batitle  \undefined \def \batitle#1{#1}\fi
\ifx \bjtitle  \undefined \def \bjtitle#1{#1}\fi
\ifx \bvolume  \undefined \def \bvolume#1{\textbf{#1}}\fi
\ifx \byear  \undefined \def \byear#1{#1}\fi
\ifx \bissue  \undefined \def \bissue#1{#1}\fi
\ifx \bfpage  \undefined \def \bfpage#1{#1}\fi
\ifx \blpage  \undefined \def \blpage #1{#1}\fi
\ifx \burl  \undefined \def \burl#1{\textsf{#1}}\fi
\ifx \doiurl  \undefined \def \doiurl#1{\url{https://doi.org/#1}}\fi
\ifx \betal  \undefined \def \betal{\textit{et al.}}\fi
\ifx \binstitute  \undefined \def \binstitute#1{#1}\fi
\ifx \binstitutionaled  \undefined \def \binstitutionaled#1{#1}\fi
\ifx \bctitle  \undefined \def \bctitle#1{#1}\fi
\ifx \beditor  \undefined \def \beditor#1{#1}\fi
\ifx \bpublisher  \undefined \def \bpublisher#1{#1}\fi
\ifx \bbtitle  \undefined \def \bbtitle#1{#1}\fi
\ifx \bedition  \undefined \def \bedition#1{#1}\fi
\ifx \bseriesno  \undefined \def \bseriesno#1{#1}\fi
\ifx \blocation  \undefined \def \blocation#1{#1}\fi
\ifx \bsertitle  \undefined \def \bsertitle#1{#1}\fi
\ifx \bsnm \undefined \def \bsnm#1{#1}\fi
\ifx \bsuffix \undefined \def \bsuffix#1{#1}\fi
\ifx \bparticle \undefined \def \bparticle#1{#1}\fi
\ifx \barticle \undefined \def \barticle#1{#1}\fi
\bibcommenthead
\ifx \bconfdate \undefined \def \bconfdate #1{#1}\fi
\ifx \botherref \undefined \def \botherref #1{#1}\fi
\ifx \url \undefined \def \url#1{\textsf{#1}}\fi
\ifx \bchapter \undefined \def \bchapter#1{#1}\fi
\ifx \bbook \undefined \def \bbook#1{#1}\fi
\ifx \bcomment \undefined \def \bcomment#1{#1}\fi
\ifx \oauthor \undefined \def \oauthor#1{#1}\fi
\ifx \citeauthoryear \undefined \def \citeauthoryear#1{#1}\fi
\ifx \endbibitem  \undefined \def \endbibitem {}\fi
\ifx \bconflocation  \undefined \def \bconflocation#1{#1}\fi
\ifx \arxivurl  \undefined \def \arxivurl#1{\textsf{#1}}\fi
\csname PreBibitemsHook\endcsname

\bibitem[\protect\citeauthoryear{Aktar et~al.}{2023}]{aktar2023predicting}
\begin{bchapter}
\bauthor{\bsnm{Aktar}, \binits{S.}},
\bauthor{\bsnm{Bärtschi}, \binits{A.}},
\bauthor{\bsnm{Badawy}, \binits{A.-H.A.}},
\bauthor{\bsnm{Oyen}, \binits{D.}},
\bauthor{\bsnm{Eidenbenz}, \binits{S.}}:
\bctitle{Predicting expressibility of parameterized quantum circuits using graph neural network}.
In: \bbtitle{2023 IEEE International Conference on Quantum Computing and Engineering (QCE)},
vol. \bseriesno{02},
pp. \bfpage{401}--\blpage{402}
(\byear{2023}).
\doiurl{10.1109/QCE57702.2023.10302}
\end{bchapter}
\endbibitem

\bibitem[\protect\citeauthoryear{Aaronson and Gottesman}{2004}]{aaronson2004improved}
\begin{barticle}
\bauthor{\bsnm{Aaronson}, \binits{S.}},
\bauthor{\bsnm{Gottesman}, \binits{D.}}:
\batitle{Improved simulation of stabilizer circuits}.
\bjtitle{Phys. Rev. A}
\bvolume{70},
\bfpage{052328}
(\byear{2004})
\doiurl{10.1103/PhysRevA.70.052328}
\end{barticle}
\endbibitem

\bibitem[\protect\citeauthoryear{Ahmadi and Greplova}{2024}]{ahmadi2024quantifying}
\begin{barticle}
\bauthor{\bsnm{Ahmadi}, \binits{A.}},
\bauthor{\bsnm{Greplova}, \binits{E.}}:
\batitle{{Quantifying non-stabilizerness via information scrambling}}.
\bjtitle{SciPost Phys.}
\bvolume{16},
\bfpage{043}
(\byear{2024})
\doiurl{10.21468/SciPostPhys.16.2.043}
\end{barticle}
\endbibitem

\bibitem[\protect\citeauthoryear{Abbas et~al.}{2021}]{abbas2021power}
\begin{barticle}
\bauthor{\bsnm{Abbas}, \binits{A.}},
\bauthor{\bsnm{Sutter}, \binits{D.}},
\bauthor{\bsnm{Zoufal}, \binits{C.}},
\bauthor{\bsnm{Lucchi}, \binits{A.}},
\bauthor{\bsnm{Figalli}, \binits{A.}},
\bauthor{\bsnm{Woerner}, \binits{S.}}:
\batitle{The power of quantum neural networks}.
\bjtitle{Nature Computational Science}
\bvolume{1}(\bissue{6}),
\bfpage{403}--\blpage{409}
(\byear{2021})
\doiurl{10.1038/s43588-021-00084-1}
\end{barticle}
\endbibitem

\bibitem[\protect\citeauthoryear{Beverland et~al.}{2020}]{beverland2020lower}
\begin{barticle}
\bauthor{\bsnm{Beverland}, \binits{M.}},
\bauthor{\bsnm{Campbell}, \binits{E.}},
\bauthor{\bsnm{Howard}, \binits{M.}},
\bauthor{\bsnm{Kliuchnikov}, \binits{V.}}:
\batitle{Lower bounds on the non-clifford resources for quantum computations}.
\bjtitle{Quantum Science and Technology}
\bvolume{5}(\bissue{3}),
\bfpage{035009}
(\byear{2020})
\doiurl{10.1088/2058-9565/ab8963}
\end{barticle}
\endbibitem

\bibitem[\protect\citeauthoryear{Bradbury et~al.}{2018}]{jax2018github}
\begin{botherref}
\oauthor{\bsnm{Bradbury}, \binits{J.}},
\oauthor{\bsnm{Frostig}, \binits{R.}},
\oauthor{\bsnm{Hawkins}, \binits{P.}},
\oauthor{\bsnm{Johnson}, \binits{M.J.}},
\oauthor{\bsnm{Leary}, \binits{C.}},
\oauthor{\bsnm{Maclaurin}, \binits{D.}},
\oauthor{\bsnm{Necula}, \binits{G.}},
\oauthor{\bsnm{Paszke}, \binits{A.}},
\oauthor{\bsnm{Vander{P}las}, \binits{J.}},
\oauthor{\bsnm{Wanderman-{M}ilne}, \binits{S.}},
\oauthor{\bsnm{Zhang}, \binits{Q.}}:
{JAX}: Composable Transformations of {P}ython+{N}um{P}y programs.
(2018).
\url{http://github.com/jax-ml/jax}
\end{botherref}
\endbibitem

\bibitem[\protect\citeauthoryear{Bergholm et~al.}{2022}]{bergholm2018pennylane}
\begin{botherref}
\oauthor{\bsnm{Bergholm}, \binits{V.}},
\oauthor{\bsnm{Izaac}, \binits{J.}},
\oauthor{\bsnm{Schuld}, \binits{M.}},
\oauthor{\bsnm{Gogolin}, \binits{C.}},
\oauthor{\bsnm{Ahmed}, \binits{S.}},
\oauthor{\bsnm{Ajith}, \binits{V.}},
\oauthor{\bsnm{Alam}, \binits{M.S.}},
\oauthor{\bsnm{Alonso-Linaje}, \binits{G.}},
\oauthor{\bsnm{AkashNarayanan}, \binits{B.}},
\oauthor{\bsnm{Asadi}, \binits{A.}}, et al.:
Pennylane: Automatic differentiation of hybrid quantum-classical computations
(2022)
{\href{https://arxiv.org/abs/1811.04968}{{arXiv:1811.04968}}}
{[quant-ph]}
\end{botherref}
\endbibitem

\bibitem[\protect\citeauthoryear{Benedetti et~al.}{2019}]{benedetti2019parameterized}
\begin{barticle}
\bauthor{\bsnm{Benedetti}, \binits{M.}},
\bauthor{\bsnm{Lloyd}, \binits{E.}},
\bauthor{\bsnm{Sack}, \binits{S.}},
\bauthor{\bsnm{Fiorentini}, \binits{M.}}:
\batitle{Parameterized quantum circuits as machine learning models}.
\bjtitle{Quantum science and technology}
\bvolume{4}(\bissue{4}),
\bfpage{043001}
(\byear{2019})
\doiurl{10.1088/2058-9565/ab4eb5}
\end{barticle}
\endbibitem

\bibitem[\protect\citeauthoryear{Bravyi and Maslov}{2021}]{bravyi2021logic}
\begin{barticle}
\bauthor{\bsnm{Bravyi}, \binits{S.}},
\bauthor{\bsnm{Maslov}, \binits{D.}}:
\batitle{Hadamard-free circuits expose the structure of the clifford group}.
\bjtitle{IEEE Transactions on Information Theory}
\bvolume{67}(\bissue{7}),
\bfpage{4546}--\blpage{4563}
(\byear{2021})
\doiurl{10.1109/TIT.2021.3081415}
\end{barticle}
\endbibitem

\bibitem[\protect\citeauthoryear{Ballarin et~al.}{2023}]{ballarin2023entanglement}
\begin{barticle}
\bauthor{\bsnm{Ballarin}, \binits{M.}},
\bauthor{\bsnm{Mangini}, \binits{S.}},
\bauthor{\bsnm{Montangero}, \binits{S.}},
\bauthor{\bsnm{Macchiavello}, \binits{C.}},
\bauthor{\bsnm{Mengoni}, \binits{R.}}:
\batitle{Entanglement entropy production in {Q}uantum {N}eural {N}etworks}.
\bjtitle{{Quantum}}
\bvolume{7},
\bfpage{1023}
(\byear{2023})
\doiurl{10.22331/q-2023-05-31-1023}
\end{barticle}
\endbibitem

\bibitem[\protect\citeauthoryear{Biamonte et~al.}{2017}]{biamonte2017quantum}
\begin{barticle}
\bauthor{\bsnm{Biamonte}, \binits{J.}},
\bauthor{\bsnm{Wittek}, \binits{P.}},
\bauthor{\bsnm{Pancotti}, \binits{N.}},
\bauthor{\bsnm{Rebentrost}, \binits{P.}},
\bauthor{\bsnm{Wiebe}, \binits{N.}},
\bauthor{\bsnm{Lloyd}, \binits{S.}}:
\batitle{Quantum machine learning}.
\bjtitle{Nature}
\bvolume{549}(\bissue{7671}),
\bfpage{195}--\blpage{202}
(\byear{2017})
\doiurl{10.1038/nature23474}
\end{barticle}
\endbibitem

\bibitem[\protect\citeauthoryear{Chen et~al.}{2024}]{chen2024magic_mps}
\begin{barticle}
\bauthor{\bsnm{Chen}, \binits{L.}},
\bauthor{\bsnm{Garcia}, \binits{R.J.}},
\bauthor{\bsnm{Bu}, \binits{K.}},
\bauthor{\bsnm{Jaffe}, \binits{A.}}:
\batitle{Magic of random matrix product states}.
\bjtitle{Phys. Rev. B}
\bvolume{109},
\bfpage{174207}
(\byear{2024})
\doiurl{10.1103/PhysRevB.109.174207}
\end{barticle}
\endbibitem

\bibitem[\protect\citeauthoryear{Ciliberto et~al.}{2018}]{ciliberto2018quantum}
\begin{barticle}
\bauthor{\bsnm{Ciliberto}, \binits{C.}},
\bauthor{\bsnm{Herbster}, \binits{M.}},
\bauthor{\bsnm{Ialongo}, \binits{A.D.}},
\bauthor{\bsnm{Pontil}, \binits{M.}},
\bauthor{\bsnm{Rocchetto}, \binits{A.}},
\bauthor{\bsnm{Severini}, \binits{S.}},
\bauthor{\bsnm{Wossnig}, \binits{L.}}:
\batitle{Quantum machine learning: a classical perspective}.
\bjtitle{Proceedings of the Royal Society A: Mathematical, Physical and Engineering Sciences}
\bvolume{474}(\bissue{2209}),
\bfpage{20170551}
(\byear{2018})
\doiurl{10.1098/rspa.2017.0551}
\end{barticle}
\endbibitem

\bibitem[\protect\citeauthoryear{{Cirq Developers}}{2025}]{cirqcode}
\begin{bbook}
\bauthor{\bsnm{{Cirq Developers}}}:
\bbtitle{Cirq}.
\bpublisher{Zenodo},
\blocation{n.p.}
(\byear{2025}).
\doiurl{10.5281/ZENODO.4062499}
\end{bbook}
\endbibitem

\bibitem[\protect\citeauthoryear{Collura et~al.}{2024}]{collura2024tensor}
\begin{bbook}
\bauthor{\bsnm{Collura}, \binits{M.}},
\bauthor{\bsnm{Lami}, \binits{G.}},
\bauthor{\bsnm{Ranabhat}, \binits{N.}},
\bauthor{\bsnm{Santini}, \binits{A.}}:
\bbtitle{Tensor Network Techniques for Quantum Computation}.
\bpublisher{SISSA Medialab s.r.l.},
\blocation{Trieste}
(\byear{2024}).
\doiurl{10.22323/9788898587049}
\end{bbook}
\endbibitem

\bibitem[\protect\citeauthoryear{Cerezo et~al.}{2022}]{cerezo22challenges}
\begin{barticle}
\bauthor{\bsnm{Cerezo}, \binits{M.}},
\bauthor{\bsnm{Verdon}, \binits{G.}},
\bauthor{\bsnm{Huang}, \binits{H.-Y.}},
\bauthor{\bsnm{Cincio}, \binits{L.}},
\bauthor{\bsnm{Coles}, \binits{P.J.}}:
\batitle{Challenges and opportunities in quantum machine learning}.
\bjtitle{Nature Computational Science}
\bvolume{2}(\bissue{9}),
\bfpage{567}--\blpage{576}
(\byear{2022})
\doiurl{10.1038/s43588-022-00311-3}
\end{barticle}
\endbibitem

\bibitem[\protect\citeauthoryear{Chen et~al.}{2021}]{chen2021onthe}
\begin{botherref}
\oauthor{\bsnm{Chen}, \binits{C.-C.}},
\oauthor{\bsnm{Watabe}, \binits{M.}},
\oauthor{\bsnm{Shiba}, \binits{K.}},
\oauthor{\bsnm{Sogabe}, \binits{M.}},
\oauthor{\bsnm{Sakamoto}, \binits{K.}},
\oauthor{\bsnm{Sogabe}, \binits{T.}}:
On the expressibility and overfitting of quantum circuit learning.
ACM Transactions on Quantum Computing
\textbf{2}(2)
(2021)
\doiurl{10.1145/3466797}
\end{botherref}
\endbibitem

\bibitem[\protect\citeauthoryear{Du et~al.}{2020}]{du2020expressive}
\begin{barticle}
\bauthor{\bsnm{Du}, \binits{Y.}},
\bauthor{\bsnm{Hsieh}, \binits{M.-H.}},
\bauthor{\bsnm{Liu}, \binits{T.}},
\bauthor{\bsnm{Tao}, \binits{D.}}:
\batitle{Expressive power of parametrized quantum circuits}.
\bjtitle{Phys. Rev. Res.}
\bvolume{2},
\bfpage{033125}
(\byear{2020})
\doiurl{10.1103/PhysRevResearch.2.033125}
\end{barticle}
\endbibitem

\bibitem[\protect\citeauthoryear{De~Mello and Ponti}{2018}]{de2018statistical}
\begin{bbook}
\bauthor{\bsnm{De~Mello}, \binits{R.F.}},
\bauthor{\bsnm{Ponti}, \binits{M.A.}}:
\bbtitle{Machine Learning: A Practical Approach on the Statistical Learning Theory},
\bedition{1st} edn.
\bpublisher{Springer},
\blocation{Cham}
(\byear{2018}).
\doiurl{10.1007/978-3-319-94989-5}
\end{bbook}
\endbibitem

\bibitem[\protect\citeauthoryear{Eisert et~al.}{2010}]{eisert2008area}
\begin{barticle}
\bauthor{\bsnm{Eisert}, \binits{J.}},
\bauthor{\bsnm{Cramer}, \binits{M.}},
\bauthor{\bsnm{Plenio}, \binits{M.B.}}:
\batitle{Colloquium: Area laws for the entanglement entropy}.
\bjtitle{Rev. Mod. Phys.}
\bvolume{82},
\bfpage{277}--\blpage{306}
(\byear{2010})
\doiurl{10.1103/RevModPhys.82.277}
\end{barticle}
\endbibitem

\bibitem[\protect\citeauthoryear{Eisert}{2013}]{eisert2013entanglement}
\begin{bchapter}
\bauthor{\bsnm{Eisert}, \binits{J.}}:
\bctitle{Entanglement and tensor network states}.
In: \beditor{\bsnm{Pavarini}, \binits{E.}},
\beditor{\bsnm{Koch}, \binits{E.}},
\beditor{\bsnm{Schollw{\"o}ck}, \binits{U.}} (eds.)
\bbtitle{Emergent Phenomena in Correlated Matter: Modeling and Simulation}
vol. \bseriesno{3}.
\bpublisher{Verlag des Forschungszentrum J{\"u}lich},
\blocation{J{\"u}lich}
(\byear{2013}).
\burl{https://www.cond-mat.de/events/correl13/manuscripts/}
\end{bchapter}
\endbibitem

\bibitem[\protect\citeauthoryear{Fan et~al.}{2025}]{fan2025disentangling}
\begin{barticle}
\bauthor{\bsnm{Fan}, \binits{C.}},
\bauthor{\bsnm{Qian}, \binits{X.}},
\bauthor{\bsnm{Zhang}, \binits{H.-C.}},
\bauthor{\bsnm{Huang}, \binits{R.-Z.}},
\bauthor{\bsnm{Qin}, \binits{M.}},
\bauthor{\bsnm{Xiang}, \binits{T.}}:
\batitle{Disentangling critical quantum spin chains with clifford circuits}.
\bjtitle{Phys. Rev. B}
\bvolume{111},
\bfpage{085121}
(\byear{2025})
\doiurl{10.1103/PhysRevB.111.085121}
\end{barticle}
\endbibitem

\bibitem[\protect\citeauthoryear{Fu et~al.}{2025}]{fu2025clifford}
\begin{barticle}
\bauthor{\bsnm{Fu}, \binits{L.}},
\bauthor{\bsnm{Shang}, \binits{H.}},
\bauthor{\bsnm{Yang}, \binits{J.}},
\bauthor{\bsnm{Guo}, \binits{C.}}:
\batitle{Clifford augmented density matrix renormalization group for ab initio quantum chemistry}.
\bjtitle{Phys. Rev. B}
\bvolume{112},
\bfpage{195111}
(\byear{2025})
\doiurl{10.1103/4ng4-vzz6}
\end{barticle}
\endbibitem

\bibitem[\protect\citeauthoryear{Frau et~al.}{2025}]{frau2025}
\begin{barticle}
\bauthor{\bsnm{Frau}, \binits{M.}},
\bauthor{\bsnm{Tarabunga}, \binits{P.S.}},
\bauthor{\bsnm{Collura}, \binits{M.}},
\bauthor{\bsnm{Tirrito}, \binits{E.}},
\bauthor{\bsnm{Dalmonte}, \binits{M.}}:
\batitle{{Stabilizer disentangling of conformal field theories}}.
\bjtitle{SciPost Phys.}
\bvolume{18},
\bfpage{165}
(\byear{2025})
\doiurl{10.21468/SciPostPhys.18.5.165}
\end{barticle}
\endbibitem

\bibitem[\protect\citeauthoryear{Gidney}{2021}]{gidney2021stim}
\begin{barticle}
\bauthor{\bsnm{Gidney}, \binits{C.}}:
\batitle{Stim: a fast stabilizer circuit simulator}.
\bjtitle{{Quantum}}
\bvolume{5},
\bfpage{497}
(\byear{2021})
\doiurl{10.22331/q-2021-07-06-497}
\end{barticle}
\endbibitem

\bibitem[\protect\citeauthoryear{Gonon and Jacquier}{2025}]{gonon2025universal}
\begin{botherref}
\oauthor{\bsnm{Gonon}, \binits{L.}},
\oauthor{\bsnm{Jacquier}, \binits{A.}}:
Universal approximation theorem and error bounds for quantum neural networks and quantum reservoirs.
IEEE Transactions on Neural Networks and Learning Systems,
1--14
(2025)
\doiurl{10.1109/TNNLS.2025.3552223}
\end{botherref}
\endbibitem

\bibitem[\protect\citeauthoryear{Gottesman}{1997}]{gottesman1997stabilizer}
\begin{bbook}
\bauthor{\bsnm{Gottesman}, \binits{D.}}:
\bbtitle{Stabilizer Codes and Quantum Error Correction}.
\bpublisher{California Institute of Technology},
\blocation{CA, USA}
(\byear{1997}).
\doiurl{10.7907/rzr7-dt72}
\end{bbook}
\endbibitem

\bibitem[\protect\citeauthoryear{Gottesman}{1998}]{gottesman1998heisenberg}
\begin{botherref}
\oauthor{\bsnm{Gottesman}, \binits{D.}}:
The heisenberg representation of quantum computers
(1998)
{\href{https://arxiv.org/abs/quant-ph/9807006}{{arXiv:quant-ph/9807006}}}
{[quant-ph]}
\end{botherref}
\endbibitem

\bibitem[\protect\citeauthoryear{Hostens et~al.}{2005}]{hostens2005stabilizer}
\begin{barticle}
\bauthor{\bsnm{Hostens}, \binits{E.}},
\bauthor{\bsnm{Dehaene}, \binits{J.}},
\bauthor{\bsnm{De~Moor}, \binits{B.}}:
\batitle{Stabilizer states and clifford operations for systems of arbitrary dimensions and modular arithmetic}.
\bjtitle{Phys. Rev. A}
\bvolume{71},
\bfpage{042315}
(\byear{2005})
\doiurl{10.1103/PhysRevA.71.042315}
\end{barticle}
\endbibitem

\bibitem[\protect\citeauthoryear{Hahn et~al.}{2022}]{hahn2022quantifying}
\begin{barticle}
\bauthor{\bsnm{Hahn}, \binits{O.}},
\bauthor{\bsnm{Ferraro}, \binits{A.}},
\bauthor{\bsnm{Hultquist}, \binits{L.}},
\bauthor{\bsnm{Ferrini}, \binits{G.}},
\bauthor{\bsnm{Garc\'{\i}a-\'Alvarez}, \binits{L.}}:
\batitle{Quantifying qubit magic resource with gottesman-kitaev-preskill encoding}.
\bjtitle{Phys. Rev. Lett.}
\bvolume{128},
\bfpage{210502}
(\byear{2022})
\doiurl{10.1103/PhysRevLett.128.210502}
\end{barticle}
\endbibitem

\bibitem[\protect\citeauthoryear{Haug and Kim}{2023}]{haug2023scalable}
\begin{barticle}
\bauthor{\bsnm{Haug}, \binits{T.}},
\bauthor{\bsnm{Kim}, \binits{M.S.}}:
\batitle{Scalable measures of magic resource for quantum computers}.
\bjtitle{PRX Quantum}
\bvolume{4},
\bfpage{010301}
(\byear{2023})
\doiurl{10.1103/PRXQuantum.4.010301}
\end{barticle}
\endbibitem

\bibitem[\protect\citeauthoryear{Harrow and Low}{2009a}]{harrow2009efficient}
\begin{bchapter}
\bauthor{\bsnm{Harrow}, \binits{A.W.}},
\bauthor{\bsnm{Low}, \binits{R.A.}}:
\bctitle{Efficient quantum tensor product expanders and k-designs}.
In: \beditor{\bsnm{Dinur}, \binits{I.}},
\beditor{\bsnm{Jansen}, \binits{K.}},
\beditor{\bsnm{Naor}, \binits{J.}},
\beditor{\bsnm{Rolim}, \binits{J.}} (eds.)
\bbtitle{Approximation, Randomization, and Combinatorial Optimization. Algorithms and Techniques},
pp. \bfpage{548}--\blpage{561}.
\bpublisher{Springer},
\blocation{Berlin, Heidelberg}
(\byear{2009}).
\doiurl{10.1007/978-3-642-03685-9\_41}
\end{bchapter}
\endbibitem

\bibitem[\protect\citeauthoryear{Harrow and Low}{2009b}]{harrow2009random}
\begin{barticle}
\bauthor{\bsnm{Harrow}, \binits{A.W.}},
\bauthor{\bsnm{Low}, \binits{R.A.}}:
\batitle{Random quantum circuits are approximate 2-designs}.
\bjtitle{Communications in Mathematical Physics}
\bvolume{291}(\bissue{1}),
\bfpage{257}--\blpage{302}
(\byear{2009})
\doiurl{10.1007/s00220-009-0873-6}
\end{barticle}
\endbibitem

\bibitem[\protect\citeauthoryear{Haug and Piroli}{2023}]{haug2023stabilizerentropies}
\begin{barticle}
\bauthor{\bsnm{Haug}, \binits{T.}},
\bauthor{\bsnm{Piroli}, \binits{L.}}:
\batitle{Stabilizer entropies and nonstabilizerness monotones}.
\bjtitle{Quantum}
\bvolume{7},
\bfpage{1092}
(\byear{2023})
\doiurl{10.22331/q-2023-08-28-1092}
\end{barticle}
\endbibitem

\bibitem[\protect\citeauthoryear{Hubregtsen et~al.}{2021}]{hubregtsen2021evaluation}
\begin{barticle}
\bauthor{\bsnm{Hubregtsen}, \binits{T.}},
\bauthor{\bsnm{Pichlmeier}, \binits{J.}},
\bauthor{\bsnm{Stecher}, \binits{P.}},
\bauthor{\bsnm{Bertels}, \binits{K.}}:
\batitle{Evaluation of parameterized quantum circuits: on the relation between classification accuracy, expressibility, and entangling capability}.
\bjtitle{Quantum Machine Intelligence}
\bvolume{3}(\bissue{1}),
\bfpage{9}
(\byear{2021})
\doiurl{10.1007/s42484-021-00038-w}
\end{barticle}
\endbibitem

\bibitem[\protect\citeauthoryear{Huang et~al.}{2025}]{huang2025augmenting}
\begin{botherref}
\oauthor{\bsnm{Huang}, \binits{J.}},
\oauthor{\bsnm{Qian}, \binits{X.}},
\oauthor{\bsnm{Li}, \binits{Z.}},
\oauthor{\bsnm{Qin}, \binits{M.}}:
Augmenting Density Matrix Renormalization Group with Matchgates and Clifford circuits
(2025).
\url{https://arxiv.org/abs/2505.08635}
\end{botherref}
\endbibitem

\bibitem[\protect\citeauthoryear{Huang et~al.}{2024}]{huang2024nonstabilizerness}
\begin{botherref}
\oauthor{\bsnm{Huang}, \binits{J.}},
\oauthor{\bsnm{Qian}, \binits{X.}},
\oauthor{\bsnm{Qin}, \binits{M.}}:
Non-stabilizerness Entanglement Entropy: a measure of hardness in the classical simulation of quantum many-body systems
(2024).
\url{https://arxiv.org/abs/2409.16895}
\end{botherref}
\endbibitem

\bibitem[\protect\citeauthoryear{Holmes et~al.}{2022}]{holmes2022connecting}
\begin{barticle}
\bauthor{\bsnm{Holmes}, \binits{Z.}},
\bauthor{\bsnm{Sharma}, \binits{K.}},
\bauthor{\bsnm{Cerezo}, \binits{M.}},
\bauthor{\bsnm{Coles}, \binits{P.J.}}:
\batitle{Connecting ansatz expressibility to gradient magnitudes and barren plateaus}.
\bjtitle{PRX Quantum}
\bvolume{3},
\bfpage{010313}
(\byear{2022})
\doiurl{10.1103/PRXQuantum.3.010313}
\end{barticle}
\endbibitem

\bibitem[\protect\citeauthoryear{Iannotti et~al.}{2025}]{iannotti2025entanglement}
\begin{botherref}
\oauthor{\bsnm{Iannotti}, \binits{D.}},
\oauthor{\bsnm{Esposito}, \binits{G.}},
\oauthor{\bsnm{Venuti}, \binits{L.C.}},
\oauthor{\bsnm{Hamma}, \binits{A.}}:
Entanglement and stabilizer entropies of random bipartite pure quantum states
(2025)
{\href{https://arxiv.org/abs/2501.19261}{{arXiv:2501.19261}}}
{[quant-ph]}
\end{botherref}
\endbibitem

\bibitem[\protect\citeauthoryear{Ibrahim et~al.}{2022}]{mohannad2022evaluation}
\begin{barticle}
\bauthor{\bsnm{Ibrahim}, \binits{M.M.}},
\bauthor{\bsnm{Mohammadbagherpoor}, \binits{H.}},
\bauthor{\bsnm{Rios}, \binits{C.}},
\bauthor{\bsnm{Bronn}, \binits{N.T.}},
\bauthor{\bsnm{Byrd}, \binits{G.T.}}:
\batitle{Evaluation of parameterized quantum circuits with cross-resonance pulse-driven entanglers}.
\bjtitle{IEEE Transactions on Quantum Engineering}
\bvolume{3},
\bfpage{1}--\blpage{13}
(\byear{2022})
\doiurl{10.1109/TQE.2022.3231124}
\end{barticle}
\endbibitem

\bibitem[\protect\citeauthoryear{Leone and Bittel}{2024}]{leone2024stabilizerentropies}
\begin{barticle}
\bauthor{\bsnm{Leone}, \binits{L.}},
\bauthor{\bsnm{Bittel}, \binits{L.}}:
\batitle{Stabilizer entropies are monotones for magic-state resource theory}.
\bjtitle{Phys. Rev. A}
\bvolume{110},
\bfpage{040403}
(\byear{2024})
\doiurl{10.1103/PhysRevA.110.L040403}
\end{barticle}
\endbibitem

\bibitem[\protect\citeauthoryear{Lami et~al.}{2025}]{lami2025quantum}
\begin{barticle}
\bauthor{\bsnm{Lami}, \binits{G.}},
\bauthor{\bsnm{Haug}, \binits{T.}},
\bauthor{\bsnm{De~Nardis}, \binits{J.}}:
\batitle{Quantum state designs with clifford-enhanced matrix product states}.
\bjtitle{PRX Quantum}
\bvolume{6},
\bfpage{010345}
(\byear{2025})
\doiurl{10.1103/PRXQuantum.6.010345}
\end{barticle}
\endbibitem

\bibitem[\protect\citeauthoryear{Liu et~al.}{2025}]{liu2025analysis}
\begin{barticle}
\bauthor{\bsnm{Liu}, \binits{Y.}},
\bauthor{\bsnm{Kaneko}, \binits{K.}},
\bauthor{\bsnm{Baba}, \binits{K.}},
\bauthor{\bsnm{Koyama}, \binits{J.}},
\bauthor{\bsnm{Kimura}, \binits{K.}},
\bauthor{\bsnm{Takeda}, \binits{N.}}:
\batitle{Analysis of parameterized quantum circuits: On the connection between expressibility and types of quantum gates}.
\bjtitle{IEEE Transactions on Quantum Engineering}
\bvolume{6},
\bfpage{1}--\blpage{12}
(\byear{2025})
\doiurl{10.1109/TQE.2025.3571484}
\end{barticle}
\endbibitem

\bibitem[\protect\citeauthoryear{Leone et~al.}{2022}]{leone2022stabilizer}
\begin{barticle}
\bauthor{\bsnm{Leone}, \binits{L.}},
\bauthor{\bsnm{Oliviero}, \binits{S.F.E.}},
\bauthor{\bsnm{Hamma}, \binits{A.}}:
\batitle{Stabilizer r\'enyi entropy}.
\bjtitle{Phys. Rev. Lett.}
\bvolume{128},
\bfpage{050402}
(\byear{2022})
\doiurl{10.1103/PhysRevLett.128.050402}
\end{barticle}
\endbibitem

\bibitem[\protect\citeauthoryear{Lancien and P{\'e}rez-Garc{\'i}a}{2022}]{lancien2022correlation}
\begin{barticle}
\bauthor{\bsnm{Lancien}, \binits{C.}},
\bauthor{\bsnm{P{\'e}rez-Garc{\'i}a}, \binits{D.}}:
\batitle{Correlation length in random mps and peps}.
\bjtitle{Annales Henri Poincar{\'e}}
\bvolume{23}(\bissue{1}),
\bfpage{141}--\blpage{222}
(\byear{2022})
\doiurl{10.1007/s00023-021-01087-4}
\end{barticle}
\endbibitem

\bibitem[\protect\citeauthoryear{Lloyd et~al.}{2020}]{lloyd2020quantum}
\begin{botherref}
\oauthor{\bsnm{Lloyd}, \binits{S.}},
\oauthor{\bsnm{Schuld}, \binits{M.}},
\oauthor{\bsnm{Ijaz}, \binits{A.}},
\oauthor{\bsnm{Izaac}, \binits{J.}},
\oauthor{\bsnm{Killoran}, \binits{N.}}:
Quantum embeddings for machine learning
(2020).
\url{https://arxiv.org/abs/2001.03622}
\end{botherref}
\endbibitem

\bibitem[\protect\citeauthoryear{Maronese et~al.}{2022}]{maronese2022quantum}
\begin{barticle}
\bauthor{\bsnm{Maronese}, \binits{M.}},
\bauthor{\bsnm{Destri}, \binits{C.}},
\bauthor{\bsnm{Prati}, \binits{E.}}:
\batitle{Quantum activation functions for quantum neural networks}.
\bjtitle{Quantum Information Processing}
\bvolume{21}(\bissue{4}),
\bfpage{128}
(\byear{2022})
\doiurl{10.1007/s11128-022-03466-0}
\end{barticle}
\endbibitem

\bibitem[\protect\citeauthoryear{Mele}{2024}]{mele2024introduction}
\begin{barticle}
\bauthor{\bsnm{Mele}, \binits{A.A.}}:
\batitle{Introduction to {H}aar measure tools in quantum information: A beginner's tutorial}.
\bjtitle{Quantum}
\bvolume{8},
\bfpage{1340}
(\byear{2024})
\doiurl{10.22331/q-2024-05-08-1340}
\end{barticle}
\endbibitem

\bibitem[\protect\citeauthoryear{Masot-Llima and Garcia-Saez}{2024}]{masot2024stabilizer}
\begin{barticle}
\bauthor{\bsnm{Masot-Llima}, \binits{S.}},
\bauthor{\bsnm{Garcia-Saez}, \binits{A.}}:
\batitle{Stabilizer tensor networks: Universal quantum simulator on a basis of stabilizer states}.
\bjtitle{Phys. Rev. Lett.}
\bvolume{133},
\bfpage{230601}
(\byear{2024})
\doiurl{10.1103/PhysRevLett.133.230601}
\end{barticle}
\endbibitem

\bibitem[\protect\citeauthoryear{Maronese and Prati}{2021}]{maronese2021continuous}
\begin{barticle}
\bauthor{\bsnm{Maronese}, \binits{M.}},
\bauthor{\bsnm{Prati}, \binits{E.}}:
\batitle{A continuous rosenblatt quantum perceptron}.
\bjtitle{International Journal of Quantum Information}
\bvolume{19}(\bissue{04}),
\bfpage{2140002}
(\byear{2021})
\doiurl{10.1142/S0219749921400025}
\end{barticle}
\endbibitem

\bibitem[\protect\citeauthoryear{Mello et~al.}{2024}]{mello2024hybrid}
\begin{barticle}
\bauthor{\bsnm{Mello}, \binits{A.F.}},
\bauthor{\bsnm{Santini}, \binits{A.}},
\bauthor{\bsnm{Collura}, \binits{M.}}:
\batitle{Hybrid stabilizer matrix product operator}.
\bjtitle{Phys. Rev. Lett.}
\bvolume{133},
\bfpage{150604}
(\byear{2024})
\doiurl{10.1103/PhysRevLett.133.150604}
\end{barticle}
\endbibitem

\bibitem[\protect\citeauthoryear{Mello et~al.}{2025}]{mello2025clifford}
\begin{barticle}
\bauthor{\bsnm{Mello}, \binits{A.F.}},
\bauthor{\bsnm{Santini}, \binits{A.}},
\bauthor{\bsnm{Lami}, \binits{G.}},
\bauthor{\bsnm{De~Nardis}, \binits{J.}},
\bauthor{\bsnm{Collura}, \binits{M.}}:
\batitle{Clifford dressed time-dependent variational principle}.
\bjtitle{Phys. Rev. Lett.}
\bvolume{134},
\bfpage{150403}
(\byear{2025})
\doiurl{10.1103/PhysRevLett.134.150403}
\end{barticle}
\endbibitem

\bibitem[\protect\citeauthoryear{Malz et~al.}{2024}]{malz2024preparation}
\begin{barticle}
\bauthor{\bsnm{Malz}, \binits{D.}},
\bauthor{\bsnm{Styliaris}, \binits{G.}},
\bauthor{\bsnm{Wei}, \binits{Z.-Y.}},
\bauthor{\bsnm{Cirac}, \binits{J.I.}}:
\batitle{Preparation of matrix product states with log-depth quantum circuits}.
\bjtitle{Phys. Rev. Lett.}
\bvolume{132},
\bfpage{040404}
(\byear{2024})
\doiurl{10.1103/PhysRevLett.132.040404}
\end{barticle}
\endbibitem

\bibitem[\protect\citeauthoryear{Nielsen and Chuang}{2010}]{nielsen2010quantum}
\begin{bbook}
\bauthor{\bsnm{Nielsen}, \binits{M.A.}},
\bauthor{\bsnm{Chuang}, \binits{I.L.}}:
\bbtitle{Quantum Computation and Quantum Information: 10th Anniversary Edition}.
\bpublisher{Cambridge University Press},
\blocation{Cambridge}
(\byear{2010}).
\doiurl{10.1017/CBO9780511976667}
\end{bbook}
\endbibitem

\bibitem[\protect\citeauthoryear{Nakaji and Yamamoto}{2021}]{nakaji2021expressibility}
\begin{barticle}
\bauthor{\bsnm{Nakaji}, \binits{K.}},
\bauthor{\bsnm{Yamamoto}, \binits{N.}}:
\batitle{Expressibility of the alternating layered ansatz for quantum computation}.
\bjtitle{{Quantum}}
\bvolume{5},
\bfpage{434}
(\byear{2021})
\doiurl{10.22331/q-2021-04-19-434}
\end{barticle}
\endbibitem

\bibitem[\protect\citeauthoryear{Qian et~al.}{2024}]{qian2024augmenting}
\begin{barticle}
\bauthor{\bsnm{Qian}, \binits{X.}},
\bauthor{\bsnm{Huang}, \binits{J.}},
\bauthor{\bsnm{Qin}, \binits{M.}}:
\batitle{Augmenting density matrix renormalization group with clifford circuits}.
\bjtitle{Phys. Rev. Lett.}
\bvolume{133},
\bfpage{190402}
(\byear{2024})
\doiurl{10.1103/PhysRevLett.133.190402}
\end{barticle}
\endbibitem

\bibitem[\protect\citeauthoryear{Qian and Qin}{2024}]{qian2024parent}
\begin{barticle}
\bauthor{\bsnm{Qian}, \binits{X.}},
\bauthor{\bsnm{Qin}, \binits{M.}}:
\batitle{Parent hamiltonian for fully augmented matrix product states}.
\bjtitle{Phys. Rev. B}
\bvolume{110},
\bfpage{121124}
(\byear{2024})
\doiurl{10.1103/PhysRevB.110.L121124}
\end{barticle}
\endbibitem

\bibitem[\protect\citeauthoryear{Ran}{2020}]{ran2020encoding}
\begin{barticle}
\bauthor{\bsnm{Ran}, \binits{S.-J.}}:
\batitle{Encoding of matrix product states into quantum circuits of one- and two-qubit gates}.
\bjtitle{Phys. Rev. A}
\bvolume{101},
\bfpage{032310}
(\byear{2020})
\doiurl{10.1103/PhysRevA.101.032310}
\end{barticle}
\endbibitem

\bibitem[\protect\citeauthoryear{Rasmussen et~al.}{2020}]{rasmussen2020reducing}
\begin{barticle}
\bauthor{\bsnm{Rasmussen}, \binits{S.E.}},
\bauthor{\bsnm{Loft}, \binits{N.J.S.}},
\bauthor{\bsnm{Bækkegaard}, \binits{T.}},
\bauthor{\bsnm{Kues}, \binits{M.}},
\bauthor{\bsnm{Zinner}, \binits{N.T.}}:
\batitle{Reducing the amount of single-qubit rotations in vqe and related algorithms}.
\bjtitle{Advanced Quantum Technologies}
\bvolume{3}(\bissue{12}),
\bfpage{2000063}
(\byear{2020})
\doiurl{10.1002/qute.202000063}
{\href{https://arxiv.org/abs/https://advanced.onlinelibrary.wiley.com/doi/pdf/10.1002/qute.202000063}{{https://advanced.onlinelibrary.wiley.com/doi/pdf/10.1002/qute.202000063}}}
\end{barticle}
\endbibitem

\bibitem[\protect\citeauthoryear{Schuld et~al.}{2020}]{schuld2020circuit}
\begin{barticle}
\bauthor{\bsnm{Schuld}, \binits{M.}},
\bauthor{\bsnm{Bocharov}, \binits{A.}},
\bauthor{\bsnm{Svore}, \binits{K.M.}},
\bauthor{\bsnm{Wiebe}, \binits{N.}}:
\batitle{Circuit-centric quantum classifiers}.
\bjtitle{Phys. Rev. A}
\bvolume{101},
\bfpage{032308}
(\byear{2020})
\doiurl{10.1103/PhysRevA.101.032308}
\end{barticle}
\endbibitem

\bibitem[\protect\citeauthoryear{Schollwöck}{2011}]{schollwoeck2011thedensity}
\begin{barticle}
\bauthor{\bsnm{Schollwöck}, \binits{U.}}:
\batitle{The density-matrix renormalization group in the age of matrix product states}.
\bjtitle{Annals of Physics}
\bvolume{326}(\bissue{1}),
\bfpage{96}--\blpage{192}
(\byear{2011})
\doiurl{10.1016/j.aop.2010.09.012} .
\bcomment{January 2011 Special Issue}
\end{barticle}
\endbibitem

\bibitem[\protect\citeauthoryear{Scott}{2015}]{scott2015multivariate}
\begin{bbook}
\bauthor{\bsnm{Scott}, \binits{D.W.}}:
\bbtitle{Multivariate Density Estimation: Theory, Practice, and Visualization}.
\bsertitle{Wiley Series in Probability and Statistics}.
\bpublisher{John Wiley \& Sons, Inc.},
\blocation{New York, Chichester}
(\byear{2015}).
\doiurl{10.1002/9781118575574} .
\bcomment{Online ISBN: 9781118575574}
\end{bbook}
\endbibitem

\bibitem[\protect\citeauthoryear{Sim et~al.}{2019}]{sim2019expressibility}
\begin{barticle}
\bauthor{\bsnm{Sim}, \binits{S.}},
\bauthor{\bsnm{Johnson}, \binits{P.D.}},
\bauthor{\bsnm{Aspuru-Guzik}, \binits{A.}}:
\batitle{Expressibility and entangling capability of parameterized quantum circuits for hybrid quantum-classical algorithms}.
\bjtitle{Advanced Quantum Technologies}
\bvolume{2}(\bissue{12}),
\bfpage{1900070}
(\byear{2019})
\doiurl{10.1002/qute.201900070}
\end{barticle}
\endbibitem

\bibitem[\protect\citeauthoryear{Schuld and Killoran}{2019}]{schuld2019quantum}
\begin{barticle}
\bauthor{\bsnm{Schuld}, \binits{M.}},
\bauthor{\bsnm{Killoran}, \binits{N.}}:
\batitle{Quantum machine learning in feature hilbert spaces}.
\bjtitle{Phys. Rev. Lett.}
\bvolume{122},
\bfpage{040504}
(\byear{2019})
\doiurl{10.1103/PhysRevLett.122.040504}
\end{barticle}
\endbibitem

\bibitem[\protect\citeauthoryear{Smith et~al.}{2024}]{smith2024constant}
\begin{barticle}
\bauthor{\bsnm{Smith}, \binits{K.C.}},
\bauthor{\bsnm{Khan}, \binits{A.}},
\bauthor{\bsnm{Clark}, \binits{B.K.}},
\bauthor{\bsnm{Girvin}, \binits{S.M.}},
\bauthor{\bsnm{Wei}, \binits{T.-C.}}:
\batitle{Constant-depth preparation of matrix product states with adaptive quantum circuits}.
\bjtitle{PRX Quantum}
\bvolume{5},
\bfpage{030344}
(\byear{2024})
\doiurl{10.1103/PRXQuantum.5.030344}
\end{barticle}
\endbibitem

\bibitem[\protect\citeauthoryear{Schuld and Petruccione}{2021}]{schuld2021machine}
\begin{bbook}
\bauthor{\bsnm{Schuld}, \binits{M.}},
\bauthor{\bsnm{Petruccione}, \binits{F.}}:
\bbtitle{Machine Learning with Quantum Computers}.
\bpublisher{Springer},
\blocation{Cham}
(\byear{2021}).
\doiurl{10.1007/978-3-030-83098-4}
\end{bbook}
\endbibitem

\bibitem[\protect\citeauthoryear{Schuld et~al.}{2021}]{schuld2021effect}
\begin{barticle}
\bauthor{\bsnm{Schuld}, \binits{M.}},
\bauthor{\bsnm{Sweke}, \binits{R.}},
\bauthor{\bsnm{Meyer}, \binits{J.J.}}:
\batitle{Effect of data encoding on the expressive power of variational quantum-machine-learning models}.
\bjtitle{Phys. Rev. A}
\bvolume{103},
\bfpage{032430}
(\byear{2021})
\doiurl{10.1103/PhysRevA.103.032430}
\end{barticle}
\endbibitem

\bibitem[\protect\citeauthoryear{Schuld et~al.}{2014}]{schuld2014quest}
\begin{barticle}
\bauthor{\bsnm{Schuld}, \binits{M.}},
\bauthor{\bsnm{Sinayskiy}, \binits{I.}},
\bauthor{\bsnm{Petruccione}, \binits{F.}}:
\batitle{The quest for a quantum neural network}.
\bjtitle{Quantum Information Processing}
\bvolume{13}(\bissue{11}),
\bfpage{2567}--\blpage{2586}
(\byear{2014})
\doiurl{10.1007/s11128-014-0809-8}
\end{barticle}
\endbibitem

\bibitem[\protect\citeauthoryear{Szombathy et~al.}{2025}]{szombathy2025independent}
\begin{botherref}
\oauthor{\bsnm{Szombathy}, \binits{D.}},
\oauthor{\bsnm{Valli}, \binits{A.}},
\oauthor{\bsnm{Moca}, \binits{C.P.}},
\oauthor{\bsnm{Farkas}, \binits{L.}},
\oauthor{\bsnm{Zaránd}, \binits{G.}}:
Independent stabilizer r\'enyi entropy and entanglement fluctuations in random unitary circuits
(2025)
{\href{https://arxiv.org/abs/2501.11489}{{arXiv:2501.11489}}}
{[quant-ph]}
\end{botherref}
\endbibitem

\bibitem[\protect\citeauthoryear{Turkeshi et~al.}{2025}]{turkeshi2025pauli}
\begin{barticle}
\bauthor{\bsnm{Turkeshi}, \binits{X.}},
\bauthor{\bsnm{Dymarsky}, \binits{A.}},
\bauthor{\bsnm{Sierant}, \binits{P.}}:
\batitle{Pauli spectrum and nonstabilizerness of typical quantum many-body states}.
\bjtitle{Phys. Rev. B}
\bvolume{111},
\bfpage{054301}
(\byear{2025})
\doiurl{10.1103/PhysRevB.111.054301}
\end{barticle}
\endbibitem

\bibitem[\protect\citeauthoryear{True and Hamma}{2022}]{true2022transitions}
\begin{barticle}
\bauthor{\bsnm{True}, \binits{S.}},
\bauthor{\bsnm{Hamma}, \binits{A.}}:
\batitle{Transitions in {E}ntanglement {C}omplexity in {R}andom {C}ircuits}.
\bjtitle{{Quantum}}
\bvolume{6},
\bfpage{818}
(\byear{2022})
\doiurl{10.22331/q-2022-09-22-818}
\end{barticle}
\endbibitem

\bibitem[\protect\citeauthoryear{Turkeshi et~al.}{2023}]{turkeshi2023measuring}
\begin{barticle}
\bauthor{\bsnm{Turkeshi}, \binits{X.}},
\bauthor{\bsnm{Schir\`o}, \binits{M.}},
\bauthor{\bsnm{Sierant}, \binits{P.}}:
\batitle{Measuring nonstabilizerness via multifractal flatness}.
\bjtitle{Phys. Rev. A}
\bvolume{108},
\bfpage{042408}
(\byear{2023})
\doiurl{10.1103/PhysRevA.108.042408}
\end{barticle}
\endbibitem

\bibitem[\protect\citeauthoryear{Vapnik}{1999}]{vapnik1999overview}
\begin{barticle}
\bauthor{\bsnm{Vapnik}, \binits{V.N.}}:
\batitle{An overview of statistical learning theory}.
\bjtitle{IEEE Transactions on Neural Networks}
\bvolume{10}(\bissue{5}),
\bfpage{988}--\blpage{999}
(\byear{1999})
\doiurl{10.1109/72.788640}
\end{barticle}
\endbibitem

\bibitem[\protect\citeauthoryear{Vapnik}{2000}]{vapnik1999nature}
\begin{bbook}
\bauthor{\bsnm{Vapnik}, \binits{V.N.}}:
\bbtitle{The Nature of Statistical Learning Theory}.
\bpublisher{Springer},
\blocation{New York, NY}
(\byear{2000}).
\doiurl{10.1007/978-1-4757-3264-1}
\end{bbook}
\endbibitem

\bibitem[\protect\citeauthoryear{Verstraete and Cirac}{2006}]{PhysRevB.73.094423}
\begin{barticle}
\bauthor{\bsnm{Verstraete}, \binits{F.}},
\bauthor{\bsnm{Cirac}, \binits{J.I.}}:
\batitle{Matrix product states represent ground states faithfully}.
\bjtitle{Phys. Rev. B}
\bvolume{73},
\bfpage{094423}
(\byear{2006})
\doiurl{10.1103/PhysRevB.73.094423}
\end{barticle}
\endbibitem

\bibitem[\protect\citeauthoryear{Veitch et~al.}{2014}]{veitch2014resource}
\begin{barticle}
\bauthor{\bsnm{Veitch}, \binits{V.}},
\bauthor{\bsnm{Hamed~Mousavian}, \binits{S.A.}},
\bauthor{\bsnm{Gottesman}, \binits{D.}},
\bauthor{\bsnm{Emerson}, \binits{J.}}:
\batitle{The resource theory of stabilizer quantum computation}.
\bjtitle{New Journal of Physics}
\bvolume{16}(\bissue{1}),
\bfpage{013009}
(\byear{2014})
\doiurl{10.1088/1367-2630/16/1/013009}
\end{barticle}
\endbibitem

\bibitem[\protect\citeauthoryear{Webb}{2016}]{webb2016theclifford}
\begin{barticle}
\bauthor{\bsnm{Webb}, \binits{Z.}}:
\batitle{The clifford group forms a unitary 3-design}.
\bjtitle{Quantum Info. Comput.}
\bvolume{16}(\bissue{15–16}),
\bfpage{1379}--\blpage{1400}
(\byear{2016})
\doiurl{10.26421/QIC16.15-16-8}
\end{barticle}
\endbibitem

\bibitem[\protect\citeauthoryear{Zhu}{2017}]{zhu2017multiqubit}
\begin{barticle}
\bauthor{\bsnm{Zhu}, \binits{H.}}:
\batitle{Multiqubit clifford groups are unitary 3-designs}.
\bjtitle{Phys. Rev. A}
\bvolume{96},
\bfpage{062336}
(\byear{2017})
\doiurl{10.1103/PhysRevA.96.062336}
\end{barticle}
\endbibitem

\bibitem[\protect\citeauthoryear{Zhang et~al.}{2025}]{zhang2025learning}
\begin{barticle}
\bauthor{\bsnm{Zhang}, \binits{F.}},
\bauthor{\bsnm{Li}, \binits{J.}},
\bauthor{\bsnm{He}, \binits{Z.}},
\bauthor{\bsnm{Situ}, \binits{H.}}:
\batitle{Learning the expressibility of quantum circuit ansatz using transformer}.
\bjtitle{Advanced Quantum Technologies}
\bvolume{8}(\bissue{6}),
\bfpage{2400366}
(\byear{2025})
\doiurl{10.1002/qute.202400366}
{\href{https://arxiv.org/abs/https://advanced.onlinelibrary.wiley.com/doi/pdf/10.1002/qute.202400366}{{https://advanced.onlinelibrary.wiley.com/doi/pdf/10.1002/qute.202400366}}}
\end{barticle}
\endbibitem

\end{thebibliography}
\end{document}